\documentclass[english,11pt, letterpaper]{article}

\usepackage{calc}
\usepackage{amsmath}
\usepackage{amssymb}
\usepackage{amsfonts}
\usepackage{graphicx}
\usepackage{bm}
\usepackage{csquotes}
\usepackage{siunitx}
\usepackage{url}
\usepackage{txfonts}

\begin{document}

\title{A vector Helmholtz electromagnetic wave propagator for inhomogeneous
  media}

\author{Laurence Keefe\footnote{National Research Council Senior Research Associate.  Current
address: Zebara LLC, Albuquerque, New
Mexico 87104},\;\;Austin McDaniel,\; Max Cubillos,\; Ilya Zilberter\footnote{National Research Council Research Associate.  Current
address: Tech-X, Boulder, Colorado 80303},\; \and Timothy Madden\\Air Force
    Research Laboratory, 3550 Aberdeen Ave SE, Kirtland AFB, New Mexico 87117,
    USA}

\date{}
  
\maketitle

\begin{abstract}
The vector electric-field Helmholtz
equation, containing  cross-polarization terms, is factored to produce
both pseudo-differential and exponential operator forms of a
three-dimensional, one-way, vector,
wave equation for propagation through inhomogeneous media. From this operator
factorization we develop a
high-order approximate, vector Helmholtz propagator that correctly
handles forward-arc,  high-angle scattering and diffraction from inhomogeneities
at all resolved length scales, and seamlessly includes evanescent waves.

Our implementation of the exponential operator form of the
one-way propagator is discussed extensively.  A rational
approximation/partial fraction decomposition of the exponential operator converts
the propagator into a moderate number of large, sparse, linear solves whose
results are summed together at each step to
advance the electric field in space. We use a new AAA-Lawson rational
interpolant for this approximation, rather than the more common Pad\'{e}
expansions that have appeared in the seismic and ocean acoustics literature previously. GMRES is used to
solve these  large systems. A direct-solve, free space propagation method
proves to be an effective preconditioner for GMRES, but can also serve as a standalone
propagator in homogeneous media.  Scalar computational
examples shown include plane wave diffraction by a circular aperture and Gaussian beam
propagation through sine-product and homogeneous refractive index fields.
The sine-product example
compares its results to that of paraxial propagation through the same media,
and demonstrates the substantial differences between these propagator
paradigms when the scale of the inhomogeneities is of the order of the
fundamental wavelength in the Helmholtz equation.  We also examine the
convergence of the homogeneous media beam results to fields generated by
Clenshaw-Curtis evaluation of the first Rayleigh-Sommerfeld integral for the same initial conditions.

\end{abstract}


\section{Introduction}

There are many computational alternatives for simulating the propagation of
electromagnetic (EM) waves, differentiated by both the mathematical character of their
formulation and the computational complexity of their implementation.  In
time-dependent problems the Finite-Difference-Time-Domain
 (FDTD) method \cite{Yee_1966} applied to (the hyperbolic system of) Maxwell's
equations is one standard. Such initial boundary value problems require fine
spatial grids for accuracy and stability, minute time integration steps to
capture transient behaviors, and the ability to implement an application-dependent range of boundary
conditions (Dirichlet, Neumann, Robin/impedance, non-reflecting). Computational resource
demands are high.  For
single-frequency/continuous-wave applications (e.g. radar cross-section
calculations),  integrations/simulations of the (elliptical)
scalar Helmholtz equation are preferred. Time integration is absent,
but the spatial gridding requirements remain, and boundary
condition versatility is still a necessity.  In addition, these are boundary value
problems that require a whole field solution, and iterative techniques are
likely necessary as grid sizes become large and direct solution techniques
become unwieldly.  When backscatter can be
neglected, and continuous-wave propagation is in the forward-arc, there are
one-way propagators, a subset of
Helmholtz methods. They are either exact or approximate scalar integral solutions (Rayleigh-Sommerfeld,
Fresnel, Fraunhofer, or their Fourier-transform equivalent Angular Spectrum
methods) valid only in homogeneous media, or
(parabolic) one-way wave equations that can be used in either homogeneous or
inhomogeneous media. One-way propagators are the simplest computationally,
because the formulations of the integrals and the one-way wave equations allow
solutions to be constructed on isolated or successive coordinate
surfaces normal to a predominant propagation direction, rather than requiring
a whole field solution over the entire
physical domain. Boundary conditions need only be applied in
two transverse dimensions, and there is no necessity for a terminating
boundary condition on the final surface of the solution.  The original one-way
wave equation in this class is the paraxial equation, derived by
Leontovich and Fock (L\&F) \cite{LeonFock1946} in 1946 for the study of long range
radio propagation near the Earth's surface.  It is a low-wavenumber
approximation of the scalar Helmholtz equation for scenarios
  where there is a distinguished/primary propagation direction.  The L\&F approximation was not
created with numerical computation in mind, but its advantages over the
Helmholtz equation in this respect were soon appreciated.  It became the
mainstay for computations of radio wave propagation both parallel and normal to the
Earth \cite{Levy2000,Knepp1983}, and spread to simulations of 
 light/laser beam propagation in turbulent
atmospheres \cite{Tatarski1961,Tatarski,Strohbehn,AndrewsPhillips},
seismic waves in the Earth \cite{Claerbout1976}, and ocean
acoustics \cite{Tappert1977}. The current work
falls into the one-way propagator class, but is neither a scalar integral solution nor a
version of the scalar paraxial equation.

The new tool we introduce here is directed towards the study of vector
electromagnetic (EM) wave propagation in
inhomogeneous media. In this paper we describe the theory and numerical procedures
underlying a one-way wave equation propagator that results from a particular
factorization of the vector electromagnetic Helmholtz equation
\begin{equation}
  \label{vec_Helm}
 \nabla ^2 \bm{E} + \nabla \left( \frac{1}{n^2} \bm{E} \cdot \nabla n^2 \right)
                              + k_0^2 n^2 \bm{E} = 0 \;.
\end{equation}
The derivation of this equation from Maxwell's equations will be recapitulated
briefly in Section 2 of this
paper, but a more thorough discussion of this general result, and the natural
appearance of the cross-polarization term, can be found in standard texts
\cite{Marcuse1972,MonYag1979,Koshiba1992,Born_Wolf1999}. Our new propagator
extends  the phenomenological domain of diffraction studies beyond that described by
the scalar three-dimensional Helmholtz equation,
and provides for the continuous coupling of electric field components throughout
the media, not just at boundaries and/or scatterers. This one-way wave equation
accurately propagates all forward-arc waves described by the Helmholtz
wavenumber dispersion relation. These include scattered and diffracted
waves at all resolved wavenumbers below the Helmholtz cutoff, $k \leq k_0n$, as well as highly damped evanescent
waves at wavenumbers beyond the Helmholtz cutoff boundary. This distinguishes it from the
paraxial equation, whose wavenumber dispersion relation is a low-wavenumber
quadratic approximation to the Helmholtz dispersion
relation, and consequently does not allow for evanescent waves.
  We expect this new tool to be particularly appropriate for
simulation of wave propagation through refractive index fields with
high-wavenumber content, as well as for initial conditions with spatial
support having wavelength or smaller scales.

Though the emphasis of
our new propagator is vector electric field propagation in inhomogeneous media, it contains a
new scalar propagator for homogeneous media, $n(x,y,z) = n_0$,  that solves
\begin{equation}
  \label{3D_Helm}
  \Delta u + (k_0 n_0)^2 u = 0
\end{equation}
via methods different from those integral techniques (Fourier transform, quadratures)
typically found in texts and the literature. Equation \eqref{3D_Helm} is one
of the  fundamental field equations of mathematical physics. It appears
naturally as an idealization of single frequency wave propagation in fluid (atmospheric and oceanic) acoustics and
seismology, and is easily derived for
electromagnetic wave propagation from Maxwell's equations.  In Cartesian
coordinates its solution
for outgoing waves in the $z\ge 0$ half-space, given a space-limited initial
condition in the $z = 0$ plane, is the first Rayleigh-Sommerfeld (RS1)
integral \cite{Goodman2005}
\begin{equation}
  \label{Ray_Som_1}
  u_{RS1}(x,y,z) = -\frac{1}{2 \pi} \iint \limits _S \frac{z}{r} u(x',y',0)
  \left( ik_0 n_0 - \frac{1}{r} \right) \frac{\exp(ik_0 n_0 r)}{r} dx' dy' \; ,
\end{equation}
where
\begin{equation}
  \label{r_define}
  r = \sqrt{(x-x')^2 + (y-y')^2 + z^2 }
\end{equation}
is the distance between the point $(x',y',0)$ in the source plane and the
observation point at $(x,y,z)$, and $S$ is the spatial region of the initial
condition.  It should be emphasized that the RS1 integral is a one-way propagator for
scalar fields in homogeneous media.

In optics \eqref{Ray_Som_1} has provided the
theoretical basis for various quadrature schemes to solve homogeneous media diffraction problems
within the plane-aperture framework of the RS1 analysis. These schemes have been
tested \cite{DubraFerrari1999,Gillen_Guha2004,Shen_Wang2006,MWhite2010,KZT2018} against the few
canonical solutions to \eqref{Ray_Som_1} that are known in the literature,
primarily the analytical expression for the field and intensity on the centerline
of a circular aperture \cite{Osterbg_Smith1961}. Reference
\cite{Lewis_beylkin_monzon2013} is a contribution to the
RS1 literature in the last decade that favors accuracy and computational
efficiency in the intermediate range between near-field and far-field over
near-field capability. It achieves this by restricting initial conditions to those
with finite support in both spatial and wavenumber domains. In so
doing it acquires the ability to calculate solutions out to ranges of a
few millions of wavelengths, a distance that none of the other RS1 techniques
have attempted. Most recently, \cite{CubillosJiminez_Ap_Num_2022a} describes a method for computing the
RS1 integral using sinc series approximation.  Once
certain quadrature weights are precomputed numerically, the accuracy of the
quadrature does not depend on wavelength  or propagation distance, only on how
well the optical field is approximated by the sinc series. In Section 5 of
the paper we will use Clenshaw-Curtis quadrature \cite{Trefethen_ATAP_ee_2020} of \eqref{Ray_Som_1} to
produce reference solutions of various beam propagation scenarios to
demonstrate the convergence behavior of our new propagator.  

The idea of extracting a one-way wave equation from factoring a
Helmholtz equation valid in inhomogeneous media comes from a framework developed for
two-dimensional scalar wave propagation in seismology and ocean acoustics
beginning more than 30 years
ago
\cite{Bamberger1etal1988,HalpernTref1988,Collins1989,CollinsssP1993,LingevitchCollins1998,Jensen2011}.
In the ocean, large
relative  variations in the refractive index cause wide-angle acoustic propagation
due to refraction, and accurate calculation of such scenarios requires more
accurate  approximations to the Helmholtz equation than that provided by the
low-wavenumber,  angle-restricted paraxial
equation. Researchers in this field developed a consistent, extendable
procedure based upon factoring a two-dimensional Helmholtz equation, then rationally
approximating the propagation operator in the resulting pseudo-differential
equation.  A two-term binomial expansion of the
pseudo-differential propagation operator yields a
paraxial equation at lowest order, but higher-order  ($\sim 20-30$) rational approximants
(Pad\'e, interpolation, least-squares, etc.) of that same operator and its
formally integrated exponential form accurately capture wave
propagation described by the appropriate Helmholtz dispersion relation at all forward-arc angles.

Researchers in EM propagation also applied this then-new simulation framework
\cite{Hadley_1992b,Schultz_etal_1994} to produce scalar one-way equations in
homogeneous media (from a factorization of \eqref{3D_Helm}), but their
low-expansion-order,  serial numerical implementations
lagged the more accurate, efficient, and parallelizable formulations developed
contemporaneously in ocean acoustics \cite{CollinsssP1993}.  These latter
methods applied to both homogeneous and inhomogeneous media, and we
extend them here to three dimensions to apply to vector EM
wave propagation  described by \eqref{vec_Helm}. 
   In our new treatment the electric
field components are coupled continuously within the three-dimensional inhomogeneous media by
the cross-polarization term in the vector Helmholtz equation, satisfy the more general displacement field
divergence condition, $\nabla \cdot \bm{\mathcal{D}} = 0$, and propagate
accurately within the entire forward arc. While our propagator method employs
non-reflecting Perfectly Matched Layer boundary conditions in the transverse
directions,  Dirichlet, Neumann, or Robin conditions can be easily
implemented.

Factoring the vector Helmholtz equation produces
both pseudo-differential and exponential operator forms of a one-way, vector,
wave equation valid in inhomogeneous media.  For the 
exponential operator form of the one-way propagator, a rational
approximation/partial fraction decomposition of the exponential operator converts
the propagator into a moderate number of large, sparse, linear solves whose
results are summed together at each step to
advance the vector electric field in space. We use a new AAA-Lawson rational
interpolant \cite{NakatsukasaTref2020} for this
approximation,  with wider validity and generally greater
accuracy than the more common Pad\'{e}
expansions that have appeared in the seismic and ocean acoustics literature previously. GMRES is used to
solve these  large systems.  Our new method for solving \eqref{3D_Helm}
proves to be an effective preconditioner for GMRES, but also serves as the default
propagator in homogeneous media.  It is independent of the integral solution
techniques usually applied to \eqref{3D_Helm}. The direct solve of these
homogeneous media preconditioner problems is simplified by the
demonstration that, in three dimensions,  they can all be put in the form of a Sylvester equation,
and solved by standard methods. Initial scalar computational
examples include Helmholtz plane wave diffraction by a circular aperture, and
Gaussian beam propagations in both homogeneous and high-wavenumber inhomogeneous
media.  The latter of these two beam simulations expose notable differences
between the results of Helmholtz and paraxial propagations.
  
The paper is organized in five additional sections.  Section 2 contains the major
theoretical results. A brief derivation 
of the vector Helmholtz equation is presented first.  Then the exponential operator and
pseudo-differential equation forms of the new EM propagator are derived by factoring
the vector Helmholtz equation, which  
highlights the importance of what will be denoted the $\bm{\bar{Z}}$ operator matrix in both the
vector Helmholtz equation and propagators. A rational approximation/partial fraction expansion of
the exponential operator form of the propagator converts it to a moderate
number of independent, large, sparse linear solves, whose results are summed at each space
step to advance the electric field in space.  Section 3 covers three specific rational
approximation ideas applicable to the decomposition of the exponential and
pseudo-differential forms of the propagator into the sparse linear solves
discussed in Section 2.
Section 4 discusses the numerical methods employed to implement the
propagator, including finite differencing, PML boundary conditions, and
preconditioning  ideas for the GMRES method applied to
the sparse linear solves.  Section 5 covers initial scalar
computational results. The first is Helmholtz diffraction by a circular aperture, where we
demonstrate the need for inital spatial resolution well beyond  the evanescent
boundary to obtain accurate results for axial values of the field and
intensity; the second shows the convergence of Gaussian beam propagations to those
generated by Clenshaw-Curtis quadrature of the Rayleigh-Sommerfeld integral;
the last is Helmholtz and paraxial propagation of beams through artificially generated
refractive index fields with high-wavenumber inhomogeneities. There we
demonstrate the substantial differences between the two propagation methods
when the inhomogeneities are on the same scale as the fundamental wavelength.
Section 6 of the paper summarizes the overall theoretical and computational results.

\section{Vector Helmholtz Equation and Two Forms of the Propagator}
	       In a lossless, chargeless medium, Maxwell's equations for the electric displacement 
	       $\bm{\mathcal{D}} = (\mathcal{D}_1 , \mathcal{D}_2 , \mathcal{D}_3)$, 
	       electric field $\bm{\mathcal{E}} = (\mathcal{E}_1 , \mathcal{E}_2 , \mathcal{E}_3)$, 
	       magnetic field $\bm{\mathcal{B}} = (\mathcal{B}_1 , \mathcal{B}_2 , \mathcal{B}_3)$,
	        and magnetizing field $\bm{\mathcal{H}} = (\mathcal{H}_1 , \mathcal{H}_2 , \mathcal{H}_3)$ are
	   \begin{subequations}
		\begin{align}
		    & \frac{\partial \bm{\mathcal{D}}}{\partial t} = \nabla \times \bm{\mathcal{H}} \\
		    \label{Faradays Law}
		    & \frac{\partial \bm{\mathcal{B}}}{\partial t} = - \nabla \times \bm{\mathcal{E}}  \\
		    \label{Gauss Law}
		    & \nabla \cdot \bm{\mathcal{D}} = 0  \\
		    & \nabla \cdot \bm{\mathcal{B}} = 0 \; .
		\end{align}
	   \end{subequations}
The constitutive relations $\bm{\mathcal{D}} = \varepsilon \bm{\mathcal{E}}$ and 
$\bm{\mathcal{B}} = \mu \bm{\mathcal{H}}$ are assumed, where $\varepsilon$ and
$\mu$ are the permittivity and permeability of the medium and
$\mu = \mu _0$, the permeability of free space.  The permittivity is typically
expressed  as $\varepsilon (\bm{r}) = \varepsilon _0 \varepsilon _r (\bm{r})$,
where $\varepsilon _0$ is the permittivity of free space and $\varepsilon _r$
is the relative  permittivity, the ratio of the permittivity of the medium to
the permittivity of free space. 

Taking the curl of \eqref{Faradays Law} the electric field $\bm{\mathcal{E}}$ satisfies
\begin{equation}
  \label{curl_wave_eq}
\nabla \times \nabla \times \bm{\mathcal{E}} 
          = - \mu _0 \varepsilon _0 \varepsilon _r (\bm{r}) \frac{\partial ^2 \bm{\mathcal{E}}}{\partial t ^2} \; .
\end{equation}
Assuming harmonic time-dependence, \(\bm{\mathcal{E}} (t, \bm{r}) = e^{i
  \omega t} \bm{E}(\bm{r}) \), and utilizing the vector identity
 \(\nabla \times \nabla \times \bm{E} = - \nabla ^2 \bm{E} + \nabla \big(
 \nabla \cdot \bm{E} \big) \),
converts \eqref{curl_wave_eq} into
\begin{equation}
\label{afterMaxwell1}
\nabla ^2 \bm{E} - \nabla \big( \nabla \cdot \bm{E} \big) 
      = - \mu _0 \varepsilon _0 \varepsilon _r (\bm{r}) \omega ^2 \bm{E} \; .
\end{equation}
From Gauss' Law \eqref{Gauss Law},
\begin{equation}
\label{from Gauss Law}
  \nabla \cdot \bm{E} = - \frac{\nabla \varepsilon}{\varepsilon} \cdot \bm{E}
  =  - \frac{\nabla n^2}{n^2} \cdot \bm{E} \; ,
\end{equation}
where \(n(\bm{r}) = \sqrt{\varepsilon_r(\bm{r})}\) is the refractive index.
Then defining the vacuum wavenumber, $k_0 = \omega \sqrt{\mu _0 \varepsilon
  _0}$ , we obtain the vector Helmholtz equation
\begin{equation}
\label{vector Helmholtz}
 \nabla ^2 \bm{E} + \nabla \left( \frac{1}{n^2} \bm{E} \cdot \nabla n^2 \right)
                              + k_0^2 n^2 \bm{E} = 0 \; .
\end{equation}
 				 				     
This vector, electric field, Helmholtz equation \eqref{vector Helmholtz} forms
the  theoretical basis of the new propagator.  The second term on the
left-hand  side describes the coupling of the components of the electric field
and the resulting cross-polarization effects.  There are many situations and
applications where it has been argued that this term is negligible and can be
dropped, thereby simplifying  \eqref{vector Helmholtz} to the scalar Helmholtz
equation.  However, for generality, we retain this term and work with the full vector
Helmholtz  equation \eqref{vector Helmholtz}.  

The propagator derivation begins with the standard assumption that, over
a  single propagation step, the refractive index field varies only in the transverse
directions, that is, $n^2 = n^2 (x,y)$.  The $z$-dependence of the refractive
index is then accounted for by modifying the refractive index field between
successive steps.  In EM wave propagation, this assumption decouples the $x, y$ electric field
components $(E_1 , E_2)$ from the $z$-component $E_3$, and plays a further crucial
role in the structure and development of the propagator.  If required, the
$E_3$ component of the field can be recovered from \eqref{Gauss Law}. An envelope assumption,
\begin{equation}
  \label{vec_envelop}
 \bm{E} (x, y, z) = \bm{w} (x, y, z) e^{i k_0 n_0 z}  \; ,
\end{equation}
with $n_0$ an arbitrary reference value of the refractive index, substituted
into
the vector Helmholtz equation \eqref{vector Helmholtz}, yields a vector
equation for the $z$-evolution of the transverse components of the envelope
electric field,  $\hat{\bm{w}} (x, y, z) = [w_1 (x, y, z), w_2 (x, y, z)]^T$, over
a single space step
\begin{equation}  \label{eq_barZode}
\frac{d^2 \bm{\hat{w}}}{dz^2} + 2 i k_0 n_0 \frac{d \bm{\hat{w}}}{dz}+ k_0^2
n_0^2 \bm{\bar{Z}} \bm{\hat{w}} = 0 \; .
\end{equation}
Here $\bar{\bm{Z}}$ is the operator matrix,
\begin{equation}
\label{Z_bar_eq}
          \bar{\bm{Z}} = \frac{1}{k_0^2 n_{0}^2} \left[ \begin{matrix}
          \Delta_{\perp} + k_0^2 (n^2 - n_{0}^2) + \frac{\partial^2 \psi}{\partial x^2} +
          \frac{\partial \psi}{\partial x} \frac{\partial}{\partial x}  &
          \frac{\partial^2 \psi}{\partial x \partial y} + \frac{\partial \psi}{\partial y}
          \frac{\partial}{\partial x} \\
          \frac{\partial^2 \psi}{\partial x \partial y} + \frac{\partial \psi}{\partial x}
          \frac{\partial}{\partial y} &
          \Delta_{\perp} + k_0^2 (n^2 - n_{0}^2) + \frac{\partial^2 \psi}{\partial y^2} +
          \frac{\partial \psi}{\partial y} \frac{\partial}{\partial y}
          \end{matrix} \right] \; ,
\end{equation}
$\psi = \ln \left( n^2 \right)$, and the transverse
Laplacian is defined by
\( \Delta_{\perp} = \frac{ \partial ^2}{\partial x^2} + \frac{\partial
  ^2}{\partial y^2}\).

Because it is assumed that the refractive index varies only in the transverse directions in a
single space step, the $z$-derivatives
in \eqref{eq_barZode} commute with the $\bar{\bm{Z}}$ operator, and the
equation is constant coefficient in $z$.  Thus it can be factored into right- and
left-going wave operators
\begin{equation}
\label{factored equation}
   \left[ \frac{d}{dz} + i k_0 n_0 \left( \bm{I} + \sqrt{\bm{I} + \bar{\bm{Z}}} \right) \right]
              \left[ \frac{d}{dz} + i k_0 n_0 \left( \bm{I} - \sqrt{\bm{I} + \bar{\bm{Z}}} \right) \right] 
                                  \hat{\bm{w}} = 0 \; .
\end{equation}
Discarding the factor corresponding to the left-going wave in \eqref{factored
  equation},  and thus explicitly neglecting backscatter, the
pseudo-differential equation form of the propagator results
\begin{equation} 
 \label{eq_barZfactor}
\left[ \frac{d}{dz} +
  i k_0 n_0 \left( \bm{I} - \sqrt{ \bm{I} + \bar{\bm{Z}}} \right) \right]
\hat{\bm{w}} = 0 \; ,
\end{equation}
which will be useful to rewrite slightly as
\begin{equation} 
 \label{eq_barZfactor2}
\frac{1}{i k_0 n_0} \frac{d}{dz} \hat{\bm{w}} + \hat{\bm{w}} = \sqrt{\bm{I} + \bar{\bm{Z}}}\ \hat{\bm{w}} \; .
\end{equation}
Equation \eqref{eq_barZfactor2} is one form of the propagator, and
historically,  it appeared
first in ocean acoustics \cite{Collins1989}, though the definition of
$\bar{\bm{Z}}$ is different in that context.  Since  
\eqref{eq_barZfactor} is constant coefficient in $z$, it can be
formally integrated \cite{CollinsssP1993} over a single space step, $\Delta z$,
to produce a second, exponential operator form of the one-way, one-step, vector propagator
\begin{equation}
\label{expop_prop}
    \hat{\bm{w}} (x,y,z + \Delta z) = e^{iK \left( -\bm{I} + \sqrt{\bm{I} + \bar{\bm{Z}} } \right) } 
                                     \hat{\bm{w}} (x,y,z) \; ,
\end{equation}
where $  K = k_0 n_0 \Delta z$ is the non-dimensional space step.
This second form of the propagator is explicitly dependent upon $K$,
but this same parameter will appear in any integration scheme developed for
the first propagator form in \eqref{eq_barZfactor2}.

In this paper we primarily discuss, and implement, the exponential operator
form of the propagator. However, there may be some accuracy advantages to be
gained using the pseudo-differential propagator, and these are considered in a
later section of the paper.  In both propagators, variation of the refractive
index in the propagation direction is implemented by updating $n(x, y)$ and
$\psi (x, y)$ in the $\bar{\bm{Z}}$ matrix for each new space step.  The size
of this space step must be at least fine enough to resolve the scale of any
inhomogeneities encountered in the $z$-direction.  However it may be finer still, if
analysis requires, and an example of such super-resolution is shown in
Section 5 of the paper.

Let us establish some notation first, and write $(a, b)$ to denote
a rational approximation of numerator degree $a$ and denominator
degree $b$. Our practical computational implementation of the exponential propagator
\eqref{expop_prop} begins with construction of an $(N,N)$ rational approximation in
 \(\bar{\bm{Z}}\) of the function consisting of the exponential of the square root:   
 \begin{equation} 
 \label{rational approximation}
   \exp{ \left( i K \sqrt{ \bm{I} + \bar{\bm{Z}}} \right) }\, \approx
   \frac{\sum_{k = 0}^{N} \alpha_{kN} \bar{\bm{Z}}^k}
        {\sum_{k = 0}^{N} \beta_{kN} \bar{\bm{Z}}^k} \; .
 \end{equation}
We then decompose this rational approximation into partial fractions
 \begin{equation} \label{pf_expan_eq}
   \frac{\sum_{k = 0}^{N} \alpha_{kN} \bar{\bm{Z}}^k}
        {\sum_{k = 0}^{N} \beta_{kN} \bar{\bm{Z}}^k} =
   \frac{\alpha_{NN}}{\beta_{NN}} + \sum_{k = 1}^{N}
   \frac{a_{kN}}{\bar{\bm{Z}} - b_{kN}\bm{I}} \; .
 \end{equation}
Employing this decomposition,  the approximate propagator can be cast in terms of
solving multiple linear systems.  Defining the $N$ auxiliary variables
\begin{equation}
  \label{Wk_define}
\overline{\bm{W}_k} = e^{-i K} \frac{a_{kN}}{\bar{\bm{Z}} - b_{kN}\bm{I}}
\hat{\bm{w}}(x,y,z) \; ,
\end{equation}
we write the approximate propagator as
 \begin{equation} \label{approx_prop}
   \hat{\bm{w}}(x,y,z+\Delta z) \approx 
  e^{-i K} \frac{\alpha_{NN}}{\beta_{NN}} \hat{\bm{w}}(x,y,z)+ \sum_{k = 1}^{N}
  \overline{\bm{W}_k} \; .
 \end{equation}
For each $k$, $\overline{\bm{W}_k}$ is found by solving the linear system 
\begin{equation}
  \label{W_k_def}
  \left( \bar{\bm{Z}} - b_{kN}\bm{I} \right) \overline{\bm{W}_k} = e^{-i K} a_{kN} \hat{\bm{w}}(x,y,z) \; .
\end{equation}
In practice, each of the $N$ versions of \eqref{W_k_def} is a large, sparse linear system, whose 
solution is obtained using GMRES.

The accuracy and range of the rational approximation to the exponential
operator is clearly an important part of this algorithm's construction.  Pad\'{e}
approximants have been the historical choice for this construction,
but we adopt a recently introduced interpolation
method \cite{NakatsukasaTref2020} that gives greater accuracy over a greater
range of \(\bar{\bm{Z}}\). Details are supplied in the next section of the
paper.

\section{Rational Approximations for Pseudo-differential and Exponential
  Operator Propagators}

An essential stage in implementing both the exponential \eqref{expop_prop} and
pseudo-differential \eqref{eq_barZfactor2} forms of the propagator is generating rational
approximations of the factor involving $\sqrt{\bm{I} + \bar{\bm{Z}}}$,  and then
decomposing that rational expression into partial fractions.  The established
procedure \cite{Collins1989,CollinsssP1993,Greene_1984} for this
construction first derives a suitably accurate scalar rational
approximation of either
$\sqrt{1 + \zeta}$ or $ \exp{ \left( i K \sqrt{ 1 + \zeta} \right) }$ in
the complex $\zeta$ plane.  Accuracy is particularly required along the real
axis, but for waveguide propagation a region above or below the real axis may
also be of
interest.  Following a
partial fraction decomposition of this rational approximation, substitution of the $\bar{\bm{Z}}$
operator for the surrogate $\zeta$ variable produces the desired expression in
\eqref{pf_expan_eq}.  An appeal to the spectral theory of operators
\cite{Greene_1984} then demonstrates that the accuracy of the spectrum of either 
 $\sqrt{\bm{I} + \bar{\bm{Z}}}$ or $\exp{ \left( i K \sqrt{ \bm{I} + \bar{\bm{Z}}}
  \right) }$ has been equally well preserved by these approximations. This
result is possibly more accessible by examining when
$\bar{\bm{Z}}$ is discretized.
For this case the relationship between the individual
eigenvalues, $\lambda_i^{\bar{\bm{Z}}} \in spec(\bar{\bm{Z}})$, and those of
the exponential operator,
$\lambda_i^{eH} \in spec\left[\exp{\left(iK (-\bm{I} +\sqrt{\bm{I} +
      \bar{\bm{Z}}}\ )\right)}\right]$,
is given by \cite{HighamFun2005}
\begin{equation}
  \label{eq_spectra_operators}
  \lambda_i^{eH} = e^{iK(-1 + \sqrt{1 + \lambda_i^{\bar{\bm{Z}}} })} \; .
\end{equation}
Equation \eqref{eq_spectra_operators} demonstrates that the accuracy of the
spectra  of the exponential operator is a direct function of the scalar
accuracy  of the original rational approximation of  $ \exp{ \left( i K \sqrt{
    1 + \zeta} \right) }$.

The range of
$\zeta$ over which the original rational approximations require
accuracy can be estimated by examining the action of the exponential operator
propagator upon the family of plane waves admitted by a single component/scalar
version of \eqref{eq_barZode} for uniform refractive index fields.  In this
case $\bar{\bm{Z}} =  \Delta_{\perp}/k_0^2 n_0^2$, and for an initial condition
of the form $w_0(x,y,0) = \exp{\left(i (k_x x + k_y y)\right)}$, the propagator is written
\begin{equation}
  \label{exp_prop_planewave}
  w(x,y,\Delta z) = \exp{\left(i K\left(-1 + \sqrt{1-\frac{k_x^2+k_y^2}{k_0^2
        n_0^2}}\right)\right)}\  w_0(x,y,0) = \exp{(i k_z \Delta z)}
w_0(x,y,0)\; ,
\end{equation}
where $k_z$ is the $z$ wavenumber calculated from the dispersion relation
obtained by substituting a plane wave solution, $w(x,y,z) = \exp{(i(k_x x+k_y
  y+k_z z))}$, into the scalar, free space version of \eqref{eq_barZode}.
The action of $\bar{\bm{Z}}$ is scalar (and pointwise) for these plane wave
initial conditions, and the argument of the exponential in the middle term in
\eqref{exp_prop_planewave}  is seen to equal
the expected $i k_z \Delta z$.  Thus, the propagator imposes the pure
phase shift experienced by the plane wave propagating over $\Delta z$.  From
this  derivation we identify $\Re(\zeta) = -\frac{k_x^2+k_y^2}{k_0^2 n_0^2}$,
where $\Re(\zeta)$ denotes the real part of $\zeta$. Thus
$\Re(\zeta) \in [-1,0]$ for propagating plane waves, and $\Re(\zeta) < -1$ for
evanescent plane waves. The remaining terms in $\bar{\bm{Z}}$
related to the refractive index field can make both positive and
negative contributions to the value of $\zeta$ within the
field.  Thus the rational expansion must deliver comparable accuracy for some positive
range of $\Re(\zeta)$.  As our default, we choose $\Re(\zeta) \in [-4,2]$ for this overall 
range, to ensure the rational approximants retain a certain formal accuracy for very wide
variations in refractive index, and for waves with wavenumbers of at least $2 k_0
n_0$,  twice those at the evanescent boundary.  This range is double that,
$\Re(\zeta) \in [-2,1]$, previously published \cite{LingevitchCollins1998} in
the ocean acoustics literature.

In all previous work we are aware of, Pad\'{e} expansions of various types were used
to initially approximate $\sqrt{1 + \zeta}$ or $\exp{ \left( i K \sqrt{ 1 +
    \zeta} \right) }$.  The use of Pad\'{e}
expansions  of  $\sqrt{1 + \zeta}$ around $\zeta = 0$ was made
popular by the description of the analytical form of this expansion's coefficients in
\cite{Bamberger1etal1988}, and the fact that for $|\zeta| < 1$ some
low-order expansions had already appeared in the literature.  However, even
at that time, there was recognition that both least-squares and interpolation
alternatives existed \cite{HalpernTref1988} that could offer superior accuracy
on the $\Re(\zeta) \in [-1,0]$ interval that characterizes
propagating wave scenarios.  The original development of the
exponential propagator described here used a Pad\'{e} approximation shifted to
$\zeta = -1$ and then rotated about that point by $\pi/8$
\cite{LingevitchCollins1998}.  The shift and rotation caused some accuracy
reduction along the $\Re(\zeta)$ axis in $[-1,0]$, but gained quite reasonable
quantitative accuracy in the evanescent region for $\Re(\zeta) < -1$. Here
we contrast the accuracy of this older approximant to a new interpolation
technique applied to both the exponential and pseudo-differential propagators, and a newly derived
Cauchy-integral approximant for the pseudo-differential propagator alone.  The new
approximants are generally much more accurate than the Pad\'{e} approximants
on the specified accuracy interval $\Re(\zeta) \in [-4, 2]$. In the case of the exponential propagator,
 the approximation maintains acceptable absolute accuracy, $\mathcal O(10^{-8})$, when the Pad\'{e}
approximant has $\mathcal O(1)$ errors for multiple-wavelength propagation steps.

\subsection{AAA-Lawson interpolation compared to Pad\'{e} expansion}

\begin{figure}[ht]
      \includegraphics[width=.7\textwidth]{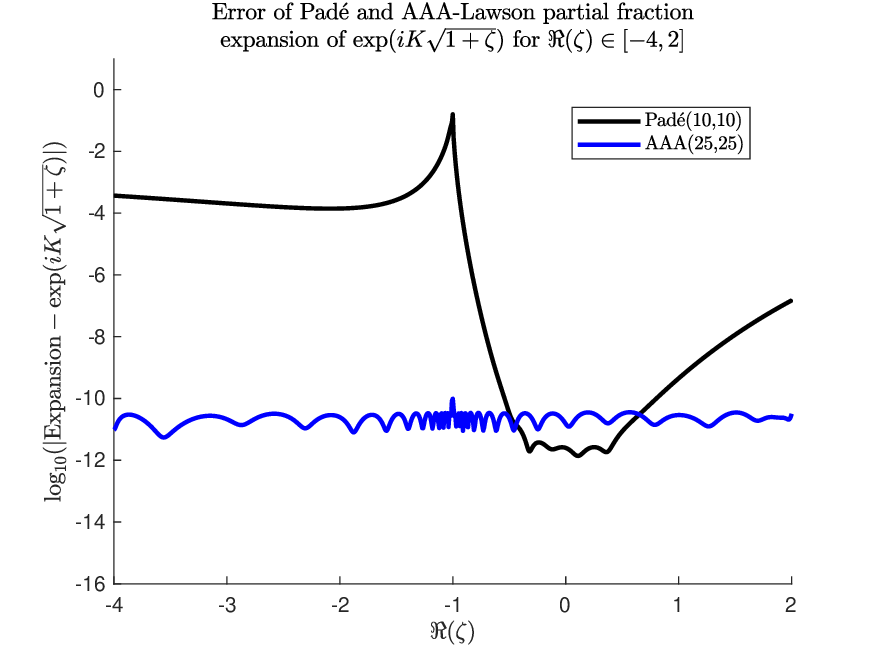}
      \caption{Error of Pad\'{e} (10,10) and AAA (25,25) partial fraction
        expansions of $ \exp{ \left( i K \sqrt{ 1 + \zeta}
            \right) }$, for  \(\Re(\zeta)\in[-4,2]\) and \(\Delta z = .5 \lambda\).}
       \label{Figure:pade_aaa_vs_expop_p5}
\end{figure}

  A new method of rational interpolation in the complex plane was recently
introduced in \cite{NakatsukasaTref2020}, and implemented in the open-source
CHEBFUN package \cite{chebfun_oxford}.  The AAA-Lawson
algorithm \enquote{first obtains a near-best
approximation and a  set of support points for a barycentric representation of
the rational approximant, then iteratively reweighted
least-squares  adjustment of the barycentric coefficients is carried out to
improve the approximation to minimax}. Customization (shifts, dilations,
contractions) of its optimization range for particular applications is easily accomplished
using the open-source software. Figure \ref{Figure:pade_aaa_vs_expop_p5} compares the approximation accuracy of
partial fraction expansions of a $(10,10)$ Pad\'{e} approximant and
$(25,25)$ AAA-Lawson interpolant to  $ \exp{ \left( i K \sqrt{ 1 + \zeta} \right) }$ on \( \Re(\zeta)
\in [-4,2] \) when $n_0 = 1.00030$ and $ \Delta z = .5 \lambda $.
The accuracy of the AAA-Lawson partial fraction interpolant is almost uniform across the \(
\Re(\zeta) \) range, over which the average error is \(\sim 2.3 \times 10^{-11}\).  The Pad\'{e} approximant
is more accurate around \(\zeta = 0\), but is substantially worse everywhere else, and particularly
degrades near the branch point of the operator at $\zeta =
-1$, which corresponds to the boundary between propagating and evanescent waves.  We note
that the $(10,10)$ Pad\'{e} approximant is at the practical limits of
computational accuracy.  Higher-order $(11,11)$ and $(12,12)$ Pad\'{e} approximants
are no more accurate than that shown, and all order Pad\'{e} expansions substantially
degrade near the branch point.
In a serial computational environment it would be justified to argue that the
increased accuracy of the AAA-Lawson interpolant has come at the cost of more than doubling the
computational effort (i.e., the need to compute 25 subsidiary solutions rather than 10). In
a parallel/MPI environment, however, this argument no longer holds, because each
term/subsidiary solution will be assigned to its own processor and compute in
parallel with all the other terms.  In practice there is very little
difference in parallel execution time when comparing these two different
expansions on the same propagation problems. 

\begin{figure}[ht]

  \includegraphics[width=.7\textwidth]{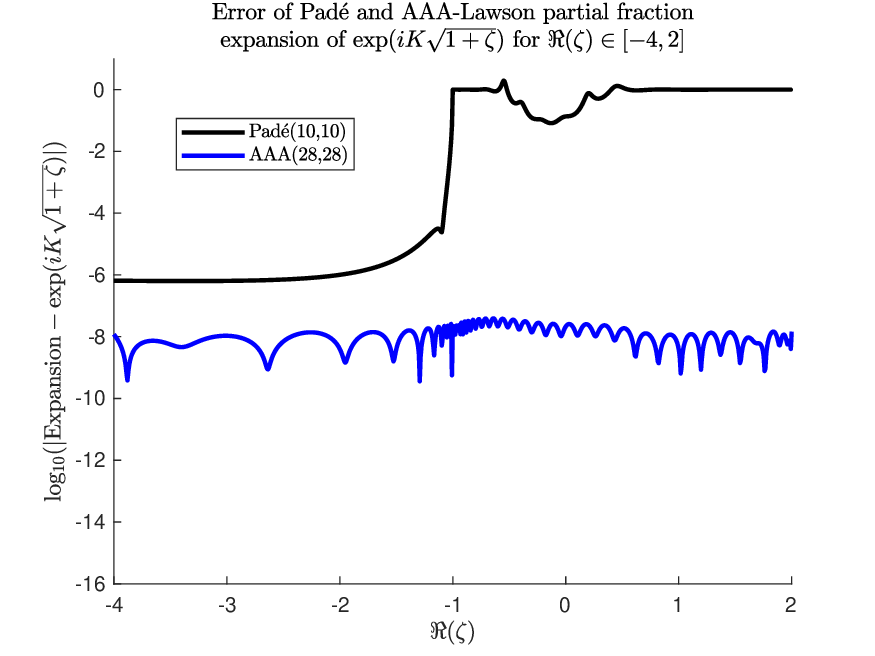}
      \caption{Error of Pad\'{e} (10,10) and AAA (28,28) partial fraction
        expansions of $ \exp{ \left( i K \sqrt{ 1 + \zeta} \right) }$, for  \(\Re(\zeta)\in[-4,2]\)
        and \(\Delta z = 5 \lambda\).}
      \label{Figure:pade_aaa_vs_expop_5}
\end{figure}

The substantial advantage of the AAA-Lawson approximation comes when
attempting to approximate the exponential propagation operator for
multiple-wavelength propagation steps.  Figure \ref{Figure:pade_aaa_vs_expop_5} compares the
approximation accuracy of partial fraction expansions of a $(10, 10)$ Pad\'{e} and
$(28, 28)$ AAA-Lawson approximation of $ \exp{ \left( i K
      \sqrt{ 1 + \zeta} \right) }$  on $ \Re(\zeta)
\in [-4,2]$ when \(n_0 = 1.00030\) and $ \Delta z = 5 \lambda $. 
 Once again, the AAA-Lawson interpolant displays almost uniform accuracy
 across the range, though errors, which now average \(\sim 1.19 \times 10^{-8}\), have
increased 3 orders of  magnitude compared to the results for the \( \Delta z =
.5 \lambda \)  approximation.  However, the Pad\'{e}
approximant is an almost complete failure in this case, displaying errors from 5 to 8
orders of magnitude greater than the AAA-Lawson interpolant across the
propagating range ($\Re(\zeta)>-1$).

    Both Pad\'{e} and AAA-Lawson approximations are in terms of \(\zeta \), but
their expansion coefficients are functions of the non-dimensional space step, $K$, defined after \eqref{expop_prop}.  Thus,
variations in the value of $K$ due to changes in $k_0 , n_0 ,$ or $\Delta z$ require that
new expansion coefficients be generated. For Pad\'{e} expansion this task
was performed by the propagator during its setup phase, using the analytical
forms (computer-algebra-generated) of the first 31 Taylor series coefficients of
the exponential operator as a function of $K$.  AAA-Lawson
interpolants are more robust to these kinds of perturbations, and so a single
lookup  table may suffice, depending upon the application.  Such a lookup table consists
of the  expansion coefficients for fixed \(n_0\) and varying values
of the product \(k_0 \Delta z\).  For EM propagation in air, or the vacuum of
space ($n(x,y,z) \in [1.00000,1.00030]$), a single lookup table using $n_0=
1.00030$ fills this role. 

\subsection{A Cauchy integral partial fraction approximation}

For the pseudo-differential equation form of the propagator
\eqref{eq_barZfactor2},   we seek a partial fraction
approximation  of the complex function 
$$f(\zeta) = (1+\zeta)^{1/2}$$
that is valid on the negative real axis and some useful portion of the positive
real axis. The branch point $\zeta=-1$ is on
the negative real axis, but there is still hope of finding a good approximation away
from this branch point.  The key step is to transform the branch point at $\zeta =
-\infty$ to $w = 1$
via the M\"obius transformation
\begin{equation} \label{eq:Mobius}
  w(\zeta) = \frac{\zeta+1+i\alpha}{\zeta+1-i\alpha} \quad \longleftrightarrow \quad \zeta(w) = -1 + i\alpha \frac{w+1}{w-1} \; .
\end{equation}
This allows the branch cut in the $w$ plane to be placed on the interval $w \in [-1,1]$.  We then represent the function
\begin{equation*}
  g(w) = f(\zeta(w)) = \sqrt{\alpha}e^{i\frac{\pi}4} (w+1)^{1/2} (w-1)^{-1/2} 
\end{equation*}
using the Cauchy integral formula
\begin{equation}
  \label{eq:Cauchy}
  g(w) = \frac{1}{2\pi i} \oint_C \frac {g(\xi)}{\xi - w} d\xi \; ,
\end{equation}
evaluated on a contour that surrounds the new branch cut,
and is closed by a circular contour at infinity centered at the origin. A
picture of this contour, and the details of the contour integration can be
found in Appendix A. The resulting integral representation of $g(w)$ is 
\begin{equation}
  g(w) = c_\alpha \left( 1 - \frac 1\pi \int_{-1}^{1} \frac 1{\eta-w} \sqrt{\frac{1+\eta}{1-\eta}} d\eta \right)  \quad c_\alpha = \sqrt{\alpha} e^{i\frac{\pi}{4}}.
\end{equation}
Noting that the integral has a Jacobi weight function, we approximate
it using an $N$ point Gauss-Jacobi quadrature rule,
\begin{equation*}
  \int_{-1}^1 f(\eta) \sqrt{\frac{1+\eta}{1-\eta}} d\eta \approx \sum_{j=1}^N \omega_j f(\eta_j) \; ,
\end{equation*}
where $\omega_j$ and $\eta_j$ are the weights and nodes, respectively, associated
with the quadrature rule. For $w$ not on the branch cut, the accuracy/convergence of the
Gauss-Jacobi quadrature  is $\mathcal O(e^{-\rho N})$, where $\rho$ is the sum
of  major and minor semiaxes of the largest ellipse of analyticity with foci
at $-1$  and $1$. This rate of exponential convergence slows as $w$ approaches the branch
cut.  For $w=-1$, the singularity is removable and the function is continuous,
but with a branch point on the interval; in this case, we expect the
convergence to  be subexponential with error $\mathcal O(N^{-1/2})$.

Inserting the quadrature into the integral expression for $g$ yields
\begin{equation*}
  g(w) \approx c_\alpha \left( 1 - \frac 1\pi \sum_{j=1}^N \frac{\omega_j}{\eta_j-w} \right).
\end{equation*}
Converting from $w$ back to $\zeta$ via the M\"obius transformation produces
\begin{equation}
  \sqrt{1+\zeta} = f(\zeta) \approx c_\alpha \left( 1 + \frac 1\pi \sum_{j=1}^N
                      \frac{\omega_j}{1-\eta_j}  \frac{1-i\alpha+\zeta}{1 + i\alpha\frac{1+\eta_j}{1-\eta_j} + \zeta} \right) .
\end{equation}

We can now use this expression to obtain an approximation for the pseudo-differential equation form of the propagator \eqref{eq_barZfactor2}.  Let 
$$\widetilde \omega_j = \frac{c_\alpha}{\pi} \frac{\omega_j}{1 - \eta_j} \hspace{15pt} \mathrm{and} \hspace{15pt}   
  \sigma_j = \frac{1+\eta_j}{1-\eta_j} \; .$$
Then the right-hand side
of \eqref{eq_barZfactor2} can be expressed as
\begin{equation} \label{prime_sqrt_exp}
  \sqrt{1 + \bar{\bm{Z}}}\ \hat{\bm{w}} \approx c_\alpha \hat{\bm{w}} +
  \sum_{j=1}^N \widetilde \omega_j \left( (1 + i\alpha\sigma_j) \bm{I} + \bar{\bm{Z}} \right)^{-1} \left( (1-i\alpha) \bm{I} + \bar{\bm{Z}} \right) \hat{\bm{w}} \; .
\end{equation}
An alternative form is obtained by evaluating the synthetic division of each
individual term in the expansion:
\begin{equation} \label{pf_sqrt_exp}
  \sqrt{1 +\bar{\bm{Z}}}\ \hat{\bm{w}} \approx \left( c_\alpha +  \sum_{j=1}^N \widetilde \omega_j \right)  \hat{\bm{w}} -
    \sum_{j=1}^N \frac{i \alpha \widetilde \omega_j (1+\sigma_j)} {\left( (1 + i\alpha\sigma_j) \bm{I} + \bar{\bm{Z}} \right)} \hat{\bm{w}} \; .
\end{equation}
The form given by \eqref{pf_sqrt_exp} is the
closest to that of the partial fraction expansion \eqref{pf_expan_eq} of the rational approximation of the exponential operator
propagator.  Using \eqref{pf_sqrt_exp} would save a single $\bar{\bm{Z}}$ multiplication in each auxiliary
problem on each propagation step compared to the form in \eqref{prime_sqrt_exp}.

\begin{figure}
	\includegraphics[width=.7\textwidth]{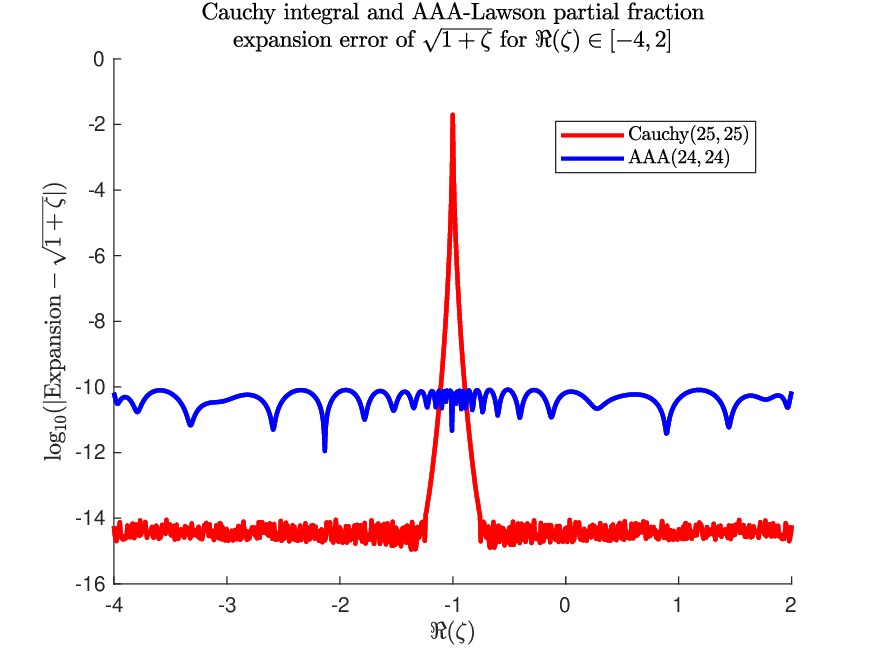}
        \caption{Error of the Cauchy integral(25,25) and AAA-Lawson(24,24) partial fraction
          expansions of $\sqrt{1+\zeta}$ for $\Re(\zeta)\in[-4,2]$.\; $\alpha = 1.0$.}
	\label{fig:Cauchy_aaa_sqrt_accuracy}
\end{figure}

The accuracy of the expansion in 
\eqref{pf_sqrt_exp} can be seen in Figure \ref{fig:Cauchy_aaa_sqrt_accuracy} confirming the
behavior we expect  from the analysis above.  Here the number of
terms is $N=25$ and the M\"obius transformation parameter is set to $\alpha =
1.0$. For comparison we also show the AAA-Lawson expansion of $\sqrt{1 +
  \zeta}$, optimized for the default $\Re(\zeta) \in [-4,2]$.  This expansion has
$N=24$ terms. Except
for a narrow band of values around the branch point at $\zeta= -1$, the Cauchy integral
expansion is noticeably better than the AAA-Lawson square-root expansion.
Note that both these expansions of the pseudo-differential operator are independent
of wavelength, refractive index, or propagator space step, in contrast to
those for the exponential operator.  
For fixed $N$, the width of the low-accuracy band around the branch point
can be shrunk substantially by setting $\alpha = 0.1$, but at the cost of reducing the
accuracy on either side of the band.  To retain the same accuracy away from
the narrowed branch point region, $N$ must be increased to 45.  Whether this
is a useful trade-off, even in an HPC computing environment, remains to be seen.

As a last step, introduce auxiliary variables $\overline{\bm{Y}_j}$ and $\overline{\bm{W}_j}$ to be the solutions of the equations
\begin{equation} \label{eq:Auxiliary_Y}
  \left( (1 + i \alpha \sigma_j) \bm{I} + \bar{\bm{Z}} \right) \overline{\bm{Y}_j} = \left( (1-i\alpha) \bm{I} + \bar{\bm{Z}} \right)\ \hat{\bm{w}}
\end{equation}
\begin{equation} \label{eq:Auxiliary_W}
  \left( (1 + i \alpha \sigma_j) \bm{I} + \bar{\bm{Z}} \right) \overline{\bm{W}_j} = -i \alpha
  \widetilde \omega_j (1+\sigma_j)\ \hat{\bm{w}} \; .
\end{equation}
Then using \eqref{prime_sqrt_exp} and \eqref{pf_sqrt_exp}, the pseudo-differential equation form of the propagator \eqref{eq_barZfactor2} becomes either
\begin{equation} \label{eq:ApproxPropagator_Y}
  \frac 1{i k_0 n_{0}} \frac{d}{dz} \hat{\bm{w}} + \hat{\bm{w}} = c_{\alpha}
  \hat{\bm{w}} + \sum_{j=1}^N \widetilde \omega_j \overline{\bm{Y}_j}
\end{equation}
or
\begin{equation} \label{eq:ApproxPropagator_W}
  \frac 1{i k_0 n_{0}} \frac{d}{dz} \hat{\bm{w}} + \hat{\bm{w}} = \left( c_{\alpha} + \sum_{j=1}^N
  \widetilde \omega_j \right) \hat{\bm{w}} + \sum_{j=1}^N \overline{\bm{W}_j} \; .
 \end{equation}
These can be implemented numerically as alternatives to the approximate
propagator \eqref{approx_prop}  coming from the rational approximation of the
exponential operator.

\section{Numerical Implementation}

\subsection{Differencing and boundary conditions}

Each of the subsidiary variables, $\overline{\bm{W}_k}$, in the approximate propagator
\eqref{approx_prop} requires a large sparse linear solve on each propagation step.
The coefficient matrices for these solves are composed from a finite difference
representation of the $(x,y)$ derivative terms in the $\bar{\bm{Z}}$
operator matrix, and updated refractive index fields, if needed.  The internal
physical  domain, gridded $m_{int} \times n_{int}$, is
surrounded by Perfectly Matched Layers (PML) of depth $m_p$ and $n_p$ respectively,
so that the total computational grid is $(m_{int}+2 m_p -1) \times (n_{int}+2
n_p-1) = m \times n$.
These wave-damping boundary conditions are applied in
the $(x,y)$ directions by modification of the peripheral entries in the second-
and first-derivative matrices that apply directly to the transverse electric
field components, $(w_1, w_2)$, across the computational
  domain. The coefficient matrices involving $n^2$ and $\psi$ use
the same differencing scheme within the
physical domain, but their entries are set to zero in the PML layers.

For wave propagation problems on a grid of resolution $\Delta x$, the resolving
efficiency  of a differencing scheme is much more important than its formal
order-of-accuracy \cite{Lele1992}.
Order-of-accuracy describes asymptotic error decrease as $\Delta x \rightarrow
0$; resolving efficiency measures the percentage of available wavenumber
bandwidth, $[0, k_{max}=\pi/\Delta x]$, over which a differentiation scheme has
no more than a specified relative error with
respect to analytical differentiation of a wave.
A wave whose derivative is calculated accurately will also
have accurate phase and group velocities.  The band of dispersive waves
correctly  propagated on a given grid is the band of waves accurately
differentiated on that grid. The differencing schemes for second and first
derivatives that our method employs are the fourth-order, explicit, 13-point
schemes described in \cite{LRKeefe2012} that were 
originally developed for application to a matrix
exponential solver of the paraxial equation.  The resolving
efficiency of the second derivative matrices at the 99.5\% relative accuracy level is
83\% of the grid bandwidth, an order of magnitude wider than that attained by standard,
3-point, second-order differencing on the same grid. The PML
regions in these differencing matrices use the complex-coordinate
transformation paradigm \cite{ChewWeedon1994,Rappaport1995,Petropoulos2000,Becache-Joly2002,BPG2004}.
Further details can be found in \cite{LRKeefe2012}. 

\subsection{GMRES iterations and preconditioning}

Reference \cite{Saad2003} (Section 9.5, page 270) describes the logical structure of a generic,
right-conditioned GMRES algorithm, but does not specify the preconditioner,
since this choice will be problem dependent. To this algorithm we join a new preconditioner based
upon directly solving a free space problem at each GMRES iteration for each
\(\overline{\bm{W}_k}\).  For a specified GMRES relative error tolerance no
greater than $\tau = 10^{-10}$, the resulting
method only requires one iteration for a homogeneous media problem ($\tau < 10^{-13}$), and rarely
more than five for inhomogeneous media ($\tau < 10^{-11}$). Its memory usage
is  minimal, since it does not compute or store
any actual matrix decompositions, such as those required by the ILU
preconditioning method \cite{Saad2003}.  In addition, the new
GMRES method uses Householder transformations to maintain orthogonalization
of the Krylov basis.  To the
extent that the relative refractive index variation becomes large and continuous within the
inhomogeneous propagation domain, more iterations may be required.  Numerical
results shown later in the paper suggest that iterations begin to increase
when there is more than 3\% continuous local variation of the underlying refractive index
field. Abrupt, isolated changes in the refractive index (e.g. beam
transmission through an air-water interface) only locally increase iteration
counts.  In the next two sections the new preconditioner's physical motivation
and actual solution technique are described.

\subsubsection{Free space propagation as the preferred preconditioner}

There are many approaches to conditioning the iterative solution of a large
linear system $\bm{A} \bm{x} = \bm{b}$.
In wave propagation problems it makes sense to search for a preconditioner amongst similar, simpler
wave propagation problems that capture the skeleton, or essence, of the more
complex propagation scenario under consideration.  In the case of both
scalar and vector Helmholtz propagation in inhomogeneous media this similar, simpler problem is that
of propagation in free space, where the refractive index field is a
constant, $n = n_0$.  With reference to the structure of the
\(\bar{\bm{Z}}\) operator given by \eqref{Z_bar_eq}, the corresponding free space operator
\(\bar{\bm{Z}}_{fs}\) is
       \begin{equation}  \label{Z_bar_fs_def}
          \bar{\bm{Z}}_{fs} = \frac{1}{k_0^2 n_0^2} \left[ \begin{matrix}
          \Delta_{\perp}   &  0
           \\
           0 &
          \Delta_{\perp}
            \end{matrix} \right]
          =
           \left[ \begin{matrix}
            \tilde{\Delta}_{\perp}   &  0
            \\
            0 &
            \tilde{\Delta}_{\perp}
            \end{matrix} \right]  \; ,     
       \end{equation}
where $\tilde{\Delta}_{\perp} = (k_0 n_0)^{-2} \Delta_{\perp}$ is the scaled
transverse Laplacian.   Note that the free space operator $\bar{\bm{Z}}_{fs}$
is characterized solely  by the diffraction effects contained in the
transverse Laplacian.  The heuristic argument is that
regardless of the complexity of the three-dimensional refractive index field,
the  short-range/one-step propagation of the electromagnetic field is characterized, to zeroth
order, by how it would propagate through free space.  This argument has some similarities to that
which underpins the Born approximation, since the now-missing parts of the full
\(\bar{\bm{Z}}\) operator, which depend upon the refractive index variations, are
regarded as a perturbation (not necessarily small) on top of the free space propagation. This
heuristic motivates the use of \(\bar{\bm{Z}}_{fs} - b_{kN}\bm{I}\) as a preconditioner
for the \(\bar{\bm{Z}} - b_{kN}\bm{I}\) coefficient matrices that appear in each of the linear systems \eqref{W_k_def} 
for the subsidiary solutions \(\overline{\bm{W}_k}\).

\subsubsection{A Sylvester equation solution of the free space problem}

Because the \(\bar{\bm{Z}}_{fs}\) matrix in \eqref{Z_bar_fs_def} is block
diagonal in the vector propagation case, the same scalar free space preconditioner
will be applied to each electric field component separately during the GMRES
iterations for each $\overline{\bm{W}_k}$.  For component $i=1,2$ free space
subsidiary fields,  \(\bm{W}_k^{fs}(i) \), on the natural \(m\times n \) geometry
of the computational field, the general form of these scalar problems is
\begin{equation}
  \label{freespace_prob}
    [ \tilde{\Delta}_{\perp} - b_{kN}] \bm{W}_k^{fs}(i) = \tilde{\bm{D}_x} \bm{W}_k^{fs}(i) +
    \bm{W}_k^{fs}(i)\tilde{\bm{D}_y}^{T} - b_{kN} \bm{W}_k^{fs}(i) = e^{-iK}
    a_{kN} \bm{w}_i \; ,
\end{equation}
where $\tilde{\bm{D}_x}$ and $\tilde{\bm{D}_y}^T$ are the scaled second derivative matrices,
including PML boundary conditions, that make up $\tilde{\Delta}_{\perp}$. 
Simple matrix algebra then establishes the identity
\begin{equation}
  \label{shift_identity}
   (\tilde{\bm{D}_x}-\alpha \bm{I}_m) \bm{W}_k^{fs}(i) +
    \bm{W}_k^{fs}(i)(\tilde{\bm{D}_y} - \beta \bm{I}_n)^{T} =  \tilde{\bm{D}_x} \bm{W}_k^{fs}(i) +
    \bm{W}_k^{fs}(i)\tilde{\bm{D}_y}^{T} - b_{kN} \bm{W}_k^{fs}(i)
\end{equation}
provided $\alpha + \beta = b_{kN}$.  The symmetric alternative, $\alpha=\beta=.5
b_{kN}$ is chosen here.
With this identity, each solution of the component subsidiary field problems in
\eqref{freespace_prob} can be obtained from a Sylvester equation
\begin{equation}
  \label{fs_sylvester}
   (\tilde{\bm{D}_x}-.5 b_{kN} \bm{I}_m) \bm{W}_k^{fs}(i) +
  \bm{W}_k^{fs}(i)(\tilde{\bm{D}_y} - .5 b_{kN} \bm{I}_n)^{T} =  e^{-iK}
  a_{kN} \bm{w}_i \; .
\end{equation}
These problems can be solved using standard numerical techniques
(Bartels \& Stewart, Hessenberg-Schur) available in LAPACK, Matlab, Slicot, NAG,
etc.  The full free space propagator then has the form
 \begin{equation} \label{approx_fs_prop}
   \bm{w}_i(x,y,z+\Delta z) \approx 
  e^{-i K} \frac{\alpha_{NN}}{\beta_{NN}} \bm{w}_i(x,y,z)+ \sum_{k = 1}^{N}
  \bm{W}_k^{fs}(i).
 \end{equation}

 These same preconditioner ideas also apply to solution methods for the subsidiary
 fields $\overline{\bm{Y}_j}$ and $\overline{\bm{W}_j}$ that appear in the 
 pseudo-differential forms of the vector propagator shown in
 \eqref{eq:ApproxPropagator_Y}  and \eqref{eq:ApproxPropagator_W}.

\section{Scalar Computational Results}
\subsection{The aperture problem and its convergence}

A useful theoretical result that tests the functionality of a Helmholtz
propagator is the analytical solution of the
RS1 integral \eqref{Ray_Som_1} due to Osterberg and Smith (O\&S)
\cite{Osterbg_Smith1961} describing the transmitted
field and  intensity on the centerline of a circular aperture illuminated by a
normal plane wave.  The problem is axisymmetric for
this configuration, and the solution reduces to a single integral over the
radial coordinate, $r'=\sqrt{x'^2+y'^2}$, in the source plane. The
centerline intensity has the functional form
\begin{equation}
  \label{eq_apersoln}
I(z) = E^2_0 \left[ 1+\frac{z^2}{z^2 + a^2} - 2 \frac{z}{\sqrt{z^2+a^2}} \cos \left(k_0
    n_0 [\sqrt{z^2+a^2} -z]\:\right)  \right] \; ,
\end{equation}
and a corresponding numerical solution of the same integral can
be obtained using standard Clenshaw-Curtis (C-C)
quadrature \cite{Trefethen_ATAP_ee_2020}. Here \(E_0\) is the incident electric
field strength in the aperture, and \(a \) is the aperture radius.

In the following we compare the RS1 analytical (O\&S) and quadrature (C-C) solutions to the
z-axis values of two Helmholtz simulations in which the $(x,y)$ grid
resolution either equals, or is less than, a wavelength.  In the
equals case, the aperture edges are not sharp on the scale
of the wavelength, and in the less-than case, they
are sharp on the scale of the wavelength.  These cases make clear that
correct solutions to problems can only occur if all relevant length scales are
resolved, and the propagation method is physically correct for those same length
scales.  Our new propagator
 naturally satisfies the propagation requirement, in both amplitude and phase, once adequate problem
 resolution has been specified.

The geometry is a circular aperture of $10 \lambda$ radius.
The physical computational domain is a $60 \lambda$ square.
Propagation  extends $120 \lambda$ in the $z$ direction,
and the propagation steps are $\lambda/20$.  This super-resolution in the
propagation direction is necessary to capture fine details of the
diffraction-mediated variation of field and intensity along the axis, despite the refractive
index field being constant.  The C-C quadratures employ the same number of
radial points for computation within the aperture as the Helmholtz propagator
uses for finite differencing within the aperture
radius, but the former are distributed on a Chebyshev grid, and the latter are
uniformly spaced.  We expect this to give some accuracy advantage to the C-C
quadratures over the propagator results.

\begin{figure}
      \includegraphics[width=.7\textwidth]{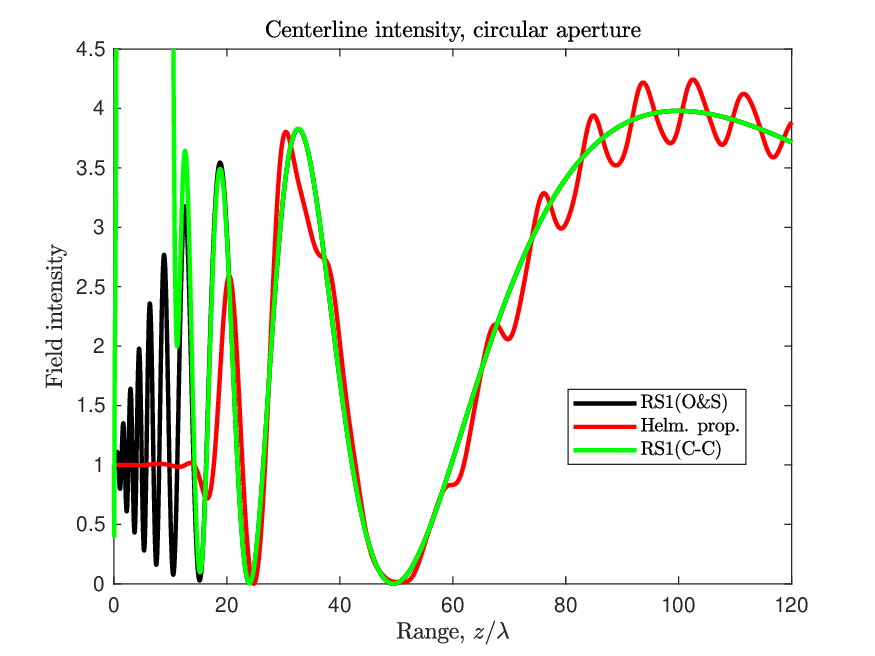}
      \caption{Centerline intensity of a plane wave diffracted by a circular
    aperture. Comparison of Rayleigh-Sommerfeld theoretical (O\&S),
    (under resolved) Rayleigh-Sommerfeld quadrature (C-C), and new (under resolved) Helmholtz propagator solutions.
    Transverse grid resolution = $\lambda$.}
      \label{fig_circ_ap_61}
\end{figure}

\begin{figure}
      \includegraphics[width=.7\textwidth]{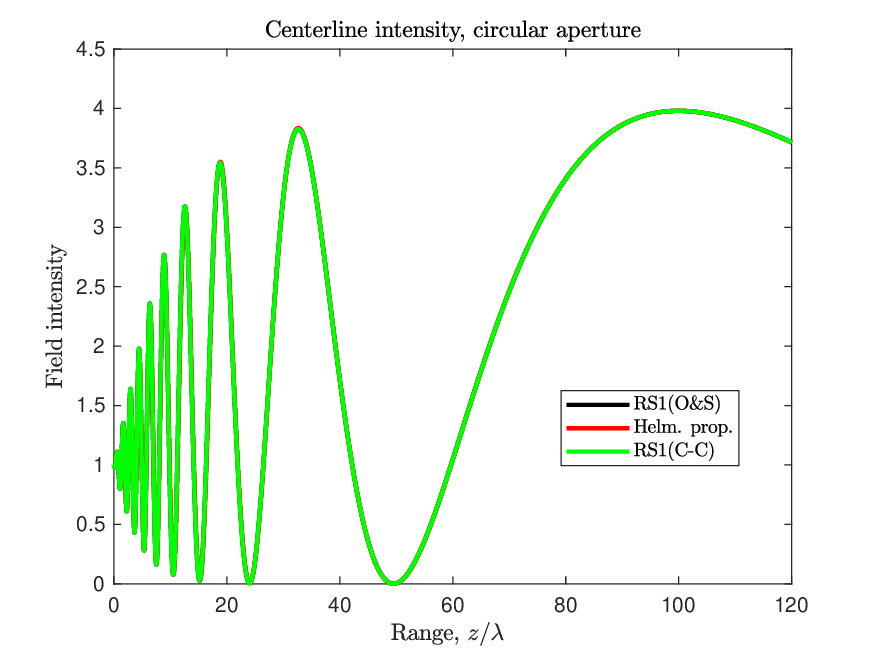}
      \caption{Centerline intensity of a plane wave diffracted by a circular
    aperture. Comparison of Rayleigh-Sommerfeld theoretical (O\&S), Rayleigh-Sommerfeld
    quadrature (C-C), and new Helmholtz propagator solutions. 
    Transverse grid resolution = $\lambda/4$.}
      \label{fig_circ_ap_241}
\end{figure}

Choosing $E_0=1$, Figures \ref{fig_circ_ap_61} and \ref{fig_circ_ap_241} display the
calculated  centerline intensity distributions for the two propagations
compared  against the analytical solution in \eqref{eq_apersoln} and the
corresponding C-C quadratures. In Figure
\ref{fig_circ_ap_61}, the physical transverse grid is $61 \times 61$, so grid
resolution is $\lambda$ and the minimum resolvable scale is $2\lambda$.  In
wavenumber terms, if $k_0 = 2\pi/\lambda$, the maximum resolved
wavenumber when $n=n_0$ is $k_0n_0/2$. PML layers of 10 grid points complete the computational
domain. In Figure \ref{fig_circ_ap_241}, the resolution is $\lambda/4$, the minimum resolvable
feature size is $\lambda/2$, and the maximum resolved wavenumber is $2 k_0
n_0$, well into the evanescent range. The physical grid is now $241 \times
241$, and the PML layers are 40 points wide.  

In Figure \ref{fig_circ_ap_61} both the less resolved simulation and the C-C
quadrature do a poor
job of capturing the analytical behavior at propagation distances
less than $20 \lambda$.  Neither capture the intensity extrema well in this
range, but the two methods fail in quite different ways.  The Helmholtz
propagator gives no extrema at all, while the C-C quadrature calculates two
peaks of magnitude greater than 200, well above the plot range.  Beyond this
initial error zone, the C-C quadrature converges to the analytical solution by
$40 \lambda$.  The propagator solution begins to follow the general trend of the
theoretical solution by $40 \lambda$, but
could not be called converged, given its continuing oscillations about the
analytical curve.  These variations do decrease with further propagation, but are
still visible at twice the distance shown. Note that they are not the result
of internal reflections, for quadrupling the depth of the PML layers neither
decreases nor eliminates them. 

The value and location
of the maxima and minima are very sensitive to the complex amplitude of the diffracted waves
that reach the centerline. The accuracy of the exponential operator expansions
shown in Figures \ref{Figure:pade_aaa_vs_expop_p5} and \ref{Figure:pade_aaa_vs_expop_5} 
guarantees that these complex amplitudes are correct for those wavenumbers that are
present.  The propagator's disagreement with the analytical Rayleigh-Sommerfeld
solution is an indication that the ensemble of waves reaching the centerline
lack the high wavenumber scales necessary to produce the theoretical interference
pattern.

Contrast these results with the more highly resolved, broader wavenumber spectrum,
simulation and quadrature in Figure \ref{fig_circ_ap_241}.  Both propagator
and quadrature show
excellent  agreement with the Rayleigh-Sommerfeld solution from aperture to
Fresnel regime, with the quadrature displaying superior accuracy for all but the
initial propagation step.  The $L2$ norms of the difference between the
calculated (propagator, C-C) intensities and the theoretical intensity in
\eqref{eq_apersoln}, normalized by the $L2$ norm of \eqref{eq_apersoln}, are
$6.58 \times 10^{-3}$ and $6.53 \times 10^{-5}$ respectively. The band of
evanescent wavenumbers in the propagator initial condition cannot
contribute  to the evolution of the field much beyond a wavelength.
However, in the very near field of  the aperture, this latter simulation shows
that their presence, correctly handled, can make the propagator solution more accurate.

\subsection{Convergence of propagator results in transverse planes to quadratures of the
  Rayleigh-Sommerfeld integral}

Analytical, closed-form solutions to the Rayleigh-Sommerfeld integral are rare,
and the aperture accuracy results just shown for both propagator simulations
and C-C quadratures were only enabled by the existence of
the solution in \eqref{eq_apersoln}. Lacking analytical solutions, additional
accuracy and convergence results with respect to other initial conditions can
only be obtained by comparison to reference solutions calculated by
alternative methods.  These may employ denser grids, larger
spatial domains, or many more terms, if the
reference comes from some sort of expansion or series.  Expected greater
accuracy is their point, not the computational efficiency of their
production. Clenshaw-Curtis quadrature of the RS1 integral showed promise as a
source of reference solutions in the last section, and is used here to create
transverse reference solutions for Gaussian beams propagated to a single
range. An important general advantage of using the RS1 integral to generate
reference solutions is that it, and its (non-trigonometric) quadratures,
do not suffer from reflection artefacts that result from truncation of an
unbounded  domain.  For the aperture this method displayed higher accuracy
than the propagator for essentially the same spatial resolution, and its
convergence behavior is quite well characterized as the number of quadrature
points  increases \cite{Trefethen_ATAP_ee_2020}.

The Gaussian beam initial conditions propagated here are created to be narrow
enough that their initial spectral content includes wavenumbers in the
evanescent range, as the apertured planewave did in the previous section. Such
narrow widths also make the beams evolve quickly in space, which decreases the
length of propagation required to cause the significant changes in their
transverse amplitude
needed for useful comparisons at a fixed range.  The Fourier-transform pair
\begin{equation}
  \label{eq_ft_gaussian}
  \frac{1}{2\pi} \int_{-\infty}^{\infty} \int_{-\infty}^{\infty}  e^{-(x^2+y^2)/w^2} e^{-i (k_x x+k_y y)} dx dy =
    \frac{w^2}{2} e^{-(k_x^2+k_y^2) w^2/4}
  \end{equation}
establishes the usual inverse relation between beam physical and spectral
widths.  We create two beams, the narrower with a (1/e) spectral width of $k_0
n_0$, and the wider with (1/e) spectral width $.5 k_0 n_0$.  The
corresponding physical widths are $w = \lambda/(\pi n_0)$ and $w=2 \lambda/(\pi
n_0)$.  Neither beam can have significant energy at wavenumbers beyond $3 k_0
n_0$,  so they are both gridded at 6-points-per-$\lambda$, in line
with basic considerations of sampling theory.  Both beams were propagated to a
distance of $.4 \lambda$ using a propagation step of $\lambda/20$. The
transverse physical domain of the wider beam was a $20\lambda$ square, gridded with
$121\times121$ points, while that of the the narrower beam was a $10\lambda$ square,
gridded with $61\times61$ points.   An additional PML grid of depth 10
surrounded both these physical domains.  In the convergence results shown
below,  the PML depths increased proportional to the grid numbers, so that the physical depth
of the PML remained constant.

The Clenshaw-Curtis reference solutions were created using the facilities of the CHEBFUN
package \cite{chebfun_oxford}. The command $RS1xy=chebfun2(F,
[A\;B\;C\;D],`equi`)$
creates a two-dimensional Chebyshev interpolant, $RS1xy$, on the
$[A,B] \times [C,D]$  domain from the $F_{ij}$ entries of matrix $F$, with the
$`equi`$  flag signaling that these values are on an equi-spaced grid.  Then
the command $sum2(RS1xy)$ returns the integral of this interpolant on the
domain.   For an $m_{int}\times n_{int}$ grid of transverse field points $(x,y)$ at
propagation distance, $z=z_p$, this two-command sequence was repeated $m_{int}n_{int}$
times to populate the propagated field values at $(x,y,z_p)$. In each
integration, the $F_{ij}$ were values of the RS1 integrand on the source plane
grid for a particular value of field point, $(x,y,z_p)$.  Starting with the
original  propagator grid, the RS1 source grid was successively refined by
factors  of 2 until the normalized Frobenius error norm between successive C-C
solutions at $z=z_p$ was $O(10^{-14})$. These error norms were based upon
differences between complex fields, not their real amplitude or intensity. This is
the native output of the RS1 integral, and a multiplication of the propagator
envelope solution by $e^{i k_0 n_0 z_p}$ (as in \eqref{vec_envelop}) yields
the same quantity.  The C-C solution at the last grid refinement was chosen as the
reference solution for that particular beam initial condition.  The use of
powers of 2 for grid refinement meant that propagations on intermediate
refined grids could all be checked for convergence to the same reference
solution  simply by suitable sampling of the reference solution.

\begin{figure}
     \includegraphics[width=.7\textwidth]{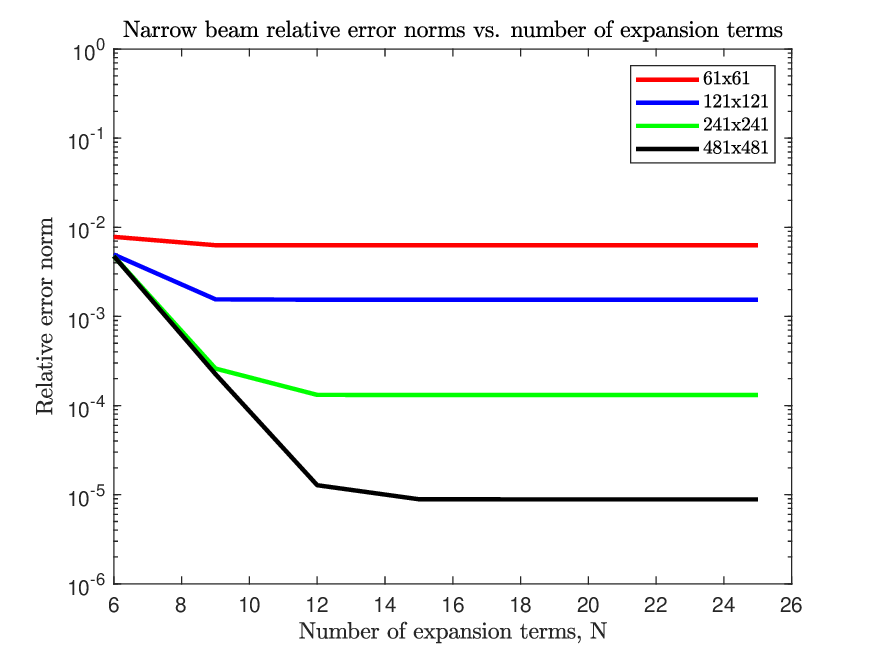}
      \caption{Normalized error norms between the Helmholtz propagated narrow beam and its
        Clenshaw-Curtis reference solution for variable number of propagator
        expansion terms and increasingly refined finite difference gridding}
    \label{var_nrw_beam_prop_rel_norms}
\end{figure}

A very simplistic
comparison between the raw execution times required by the C-C quadratures and the
propagator to obtain their beam solutions at $z_p = .4\lambda$ shows that the
propagator executes between 15 to 45 times faster than the quadratures, with
the speed advantage of the propagator increasing as the grid refinement increased.
However, this comparison does not take into  account that neither the
quadratures nor the propagator were optimized or parallelized.  The MATLAB version
of the propagator used for these comparisons was serial, used 25 expansion terms (for
optimum accuracy), and
took 8 steps to get to $z_p$,  while the quadrature calculation did this in one
step.  A subsequent propagator calculation on the narrow beam showed that its convergence
data shown in Figure \ref{var_nrw_beam_prop_rel_norms} were barely changed by the use of a single $\Delta z =
.4\lambda$ integration step.  Thus in a strictly per-step comparison, the
propagator would enjoy a further multiplicative speed advantage factor of $200 = 8 \times 25$
over the quadrature when executing in its parallelized version.  On the other hand, C-C quadratures are not the only
choice for integration of the RS1 integral
(e.g. \cite{DubraFerrari1999,Gillen_Guha2004,Shen_Wang2006,Lewis_beylkin_monzon2013,CubillosJiminez_Ap_Num_2022a}),
but there has been no systematic comparison between optimized and parallelized
versions of these methods to understand their practical accuracy and
computational requirements.  Certainly some decrease of execution time should be
expected from optimization and parallelization of any RS1 quadrature, but it seems unlikely to be of a magnitude
that would bring quadratures to execution-time parity with the propagator.
It should also be remembered that RS1 quadratures are only valid for wave propagation in constant
refractive index media, not the inhomogeneous media for which the vector
propagator has been developed.
RS1 quadratures may have advantages for computation of an isolated diffraction
field at a single range, but for any application where the development of the
intermediate wave field is of interest, they are not competitive with the
propagator method in either homogeneous or inhomogeneous media.  In addition,
the per-step comparison above demonstrates that
they are not a useful substitute for our new direct-solve homogeneous media
propagator as a preconditioner to GMRES.

\begin{figure}
     \includegraphics[width=.7\textwidth]{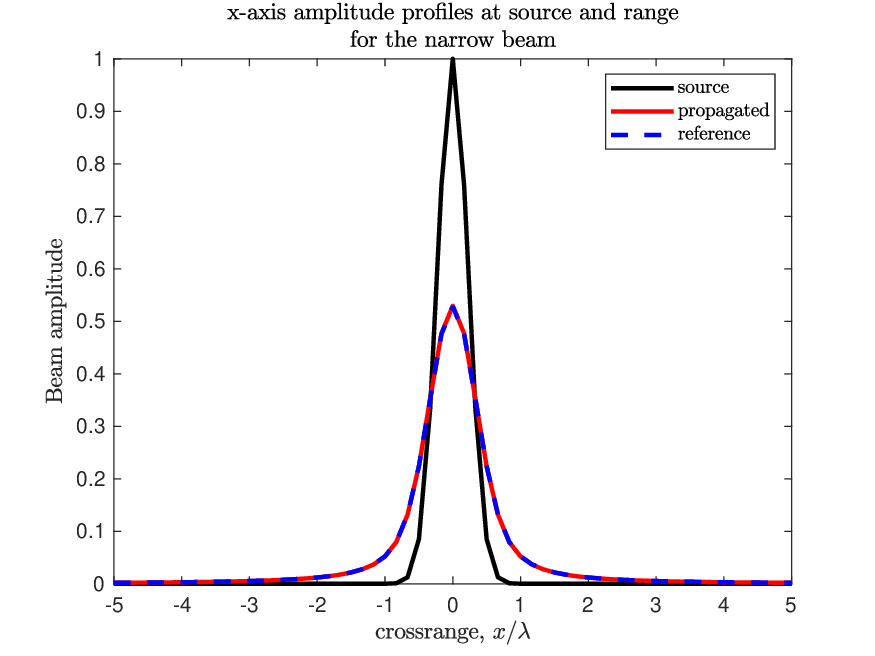}
      \caption{x-axis amplitude profiles of narrow beam at source and
        $z=.4\lambda$. Comparison of new Helmholtz propagator and Clenshaw-Curtis
    Rayleigh-Sommerfeld reference solution. Transverse grid $61\times 61$ on
    $[-5\lambda,5\lambda]$}
    \label{nrw_beam_prop_61x61}
\end{figure}

For the original source gridding, Figure \ref{nrw_beam_prop_61x61} shows the initial and $z=.4\lambda$
$x$-axis  amplitude profiles of the propagated narrow beam, along with that of
its sampled reference solution at the same range.  The propagator uses the optimum 25
term expansion of the exponential operator in this solution.
To plot resolution, the propagated field and reference solution are
indistinguishable.  The peak amplitude of the propagated solution is
$0.529509$, while that of the reference solution is $0.527961$.  In a similar
comparison of the wide beam amplitudes, the propagator and reference solutions are
similarly indistinguisable, but the peak amplitudes are now $0.920729$ and
$0.920087$.

\begin{figure}
     \includegraphics[width=.7\textwidth]{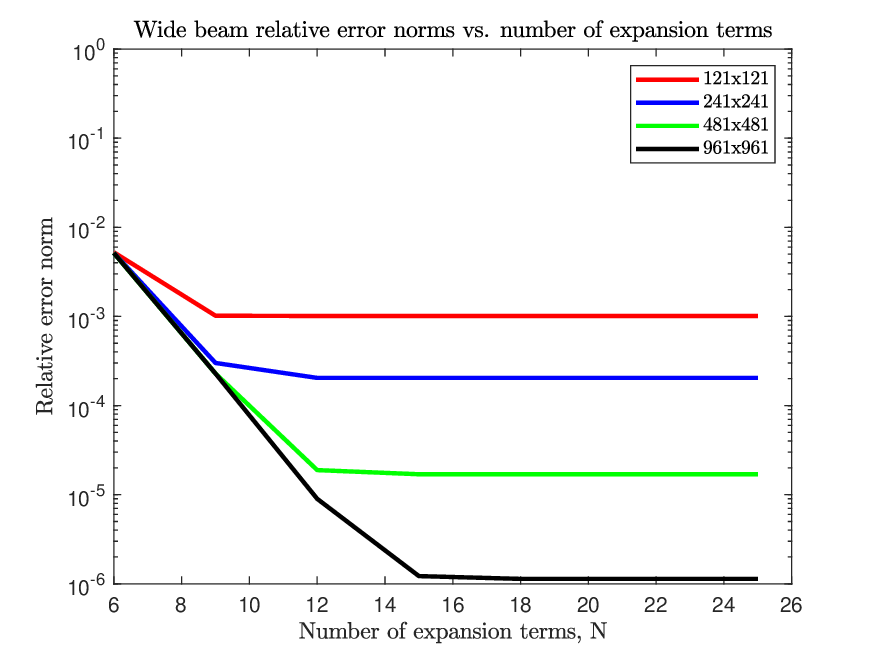}
      \caption{Normalized error norms between the Helmholtz propagated wide beam and its
        Clenshaw-Curtis reference solution for variable number of propagator
        expansion terms and increasingly refined finite difference gridding}
    \label{var_wide_beam_prop_rel_norms}
\end{figure}

Figures \ref{var_nrw_beam_prop_rel_norms} and
\ref{var_wide_beam_prop_rel_norms}  plot the Frobenius difference
norm between the Helmholtz propagated beams and their sampled reference solutions,
normalized by the Frobenius norm of the sampled reference solution.  These relative
error norms are shown as a function of both the number of exponential-operator
expansion terms and the finite differencing grid employed by the propagator.

Notable for both beams is the insensitivity of the error norms to the number of
expansion terms in the propagator once 18 terms are utilized, and this is
true regardless of finite differencing resolution.  For the same increase in
number of expansion terms as shown in the plots, the error norm of the
exponential operator expansion for this propagation step decreases by seven
orders of magnitude. This suggests that the
exponential operator expansion is the most accurate part of the overall
numerical method, and that global accuracy is controlled by the finite differencing/discretization
error, with possibly some contributions from the PML.  The results certainly
suggest that fewer than the optimium number of expansion terms could be used
in practical computations.  This might be advantageous, since fewer processors
and less memory would be required. However, any expectations of a decrease in
wall-clock execution time in a parallel computing environment must be tempered
by the knowledge that each propagation step cannot occur until the processor(s)
calculating the term requiring the most GMRES iterations completes its task.
For homogeneous media problems there will be no speed-up, since all terms
require a single iteration.  For inhomogeneous media propagations, the latter
expansion terms typically do require more GMRES iterations per propagation
step, but this number must certainly be dependent upon the strength and kind
of the inhomogeneities in ways that are difficult to quantify.  Maximum GMRES
iterations experienced so far are typically 5-8, so it seems unlikely that
more than a 20\% execution time decrease could be obtained by eliminating
calculations of these latter terms in a parallel computing environment.

\subsection{Beam propagation through high-wavenumber inhomogeneities}

The paraxial equation has been a widely utilized model for EM
wave propagation in weakly inhomogeneous media
since its introduction in 1946 \cite{LeonFock1946,Levy2000}.  Here we contrast
beam propagation results from that equation with those from our new Helmholtz
propagator in artificially generated refractive index fields.  The form of
these three-dimensional refractive index fields is that of a rotated sine-product:
\begin{equation}
  \label{eq_sineprod}
  n(x,y,z) = n_0 + \Delta n\,  \sin(k_s \tilde{x}) \sin(k_s \tilde{y})
  \sin(k_s \tilde{z}) \; ,
\end{equation}
where $k_s= k_0 n_0/(\sqrt{3} q),\: q \in \{.5,1,2 \}$, and the rotated tilde
system is related to the propagator coordinate system through the axis-angle
formulation \cite{Palazzolo_1976}.  In the propagator system the (randomly
chosen) rotation axis and angle were, respectively, the unit vector $\hat{u} = (0.258,0.312,0.914)$,
and the angle 0.128 radians.  The selected values of $q$ yield refractive index
fields with fundamental wavenumbers of $2 k_0 n_0,\, k_0 n_0$, and $.5 k_0
n_0$.  The reference refractive index value was $n_0 = 1.2$, and the amplitude
of the refractive index variations was $\Delta n = 0.025$.

\begin{figure}[ht]
      \includegraphics[width=.8\textwidth]{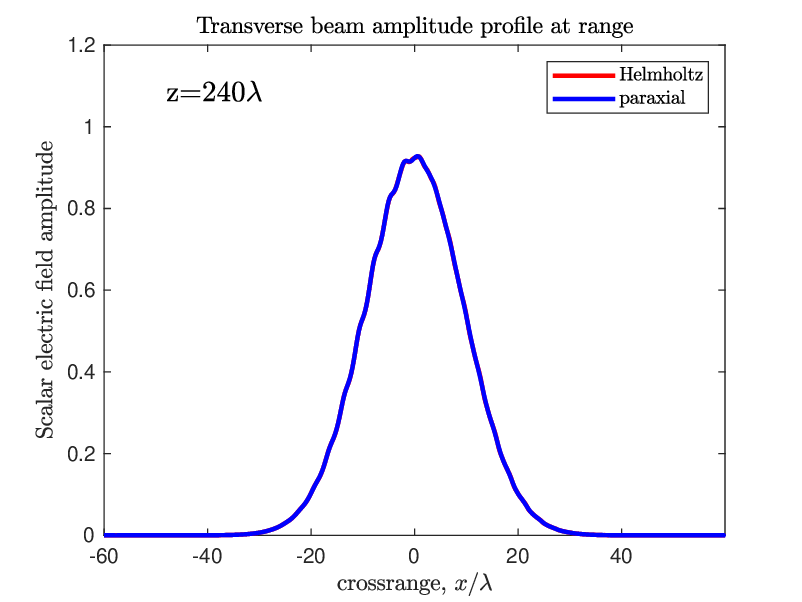}
      \caption{Beam amplitude profiles on the x-axis after $240 \lambda$ Helmholtz and
        paraxial propagations through a sine-product refractive index field
        with a $.5 k_0 n_0$ fundamental wavenumber}
      \label{Figure:sine_prod_n_xamp_20mu}
\end{figure}

\begin{figure}[ht]
      \includegraphics[width=.7\textwidth]{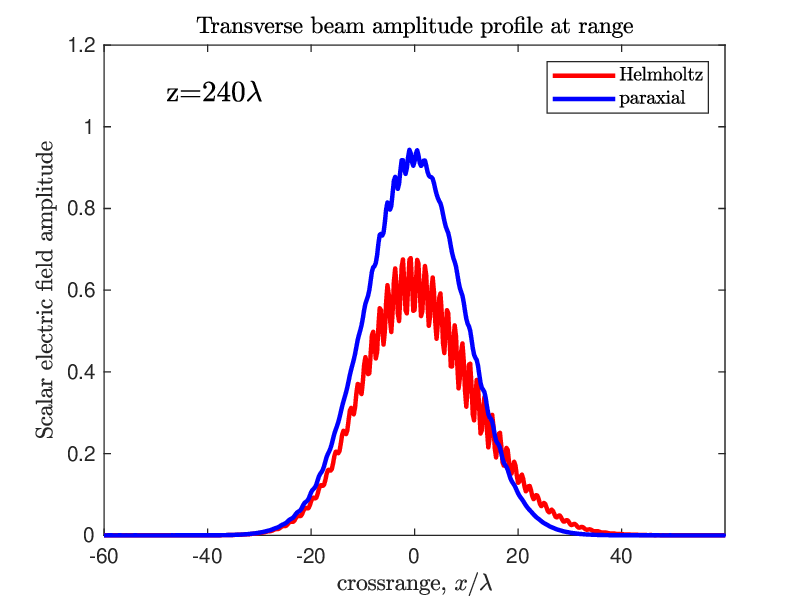}
      \caption{Beam amplitude profiles on the x-axis after $240 \lambda$ Helmholtz and
        paraxial propagations through a sine-product refractive index field
        with a $k_0 n_0$ fundamental wavenumber}
       \label{Figure:sine_prod_n_xamp_10mu}
\end{figure}

\begin{figure}[ht]
      \includegraphics[width=.7\textwidth]{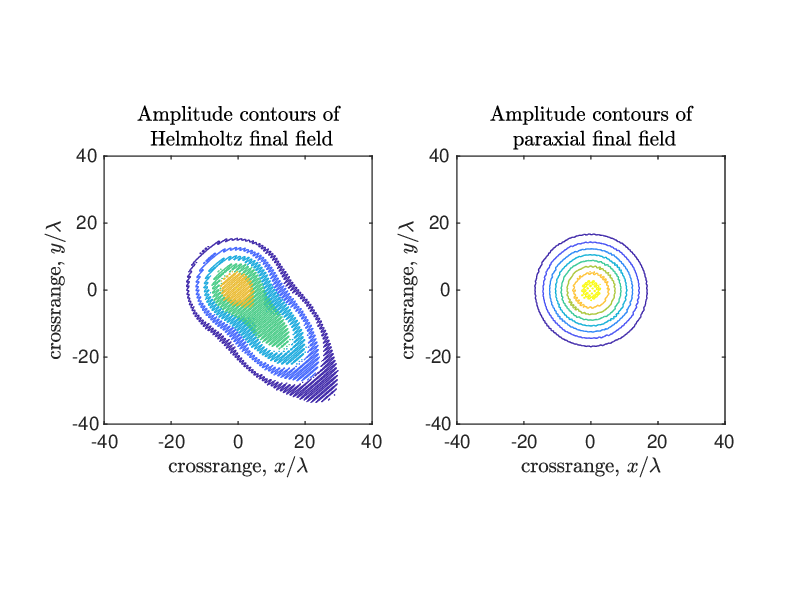}
      \caption{Transverse amplitude contours of Helmholtz and paraxial beams
        after a $z=240 \lambda$ propagation through a sine-product refractive
        index field with a $k_0 n_0$ fundamental wavelength}
      \label{Figure:sine_prod_n_amp_contour_10mu}
\end{figure}

\begin{figure}[ht]
      \includegraphics[width=.7\textwidth]{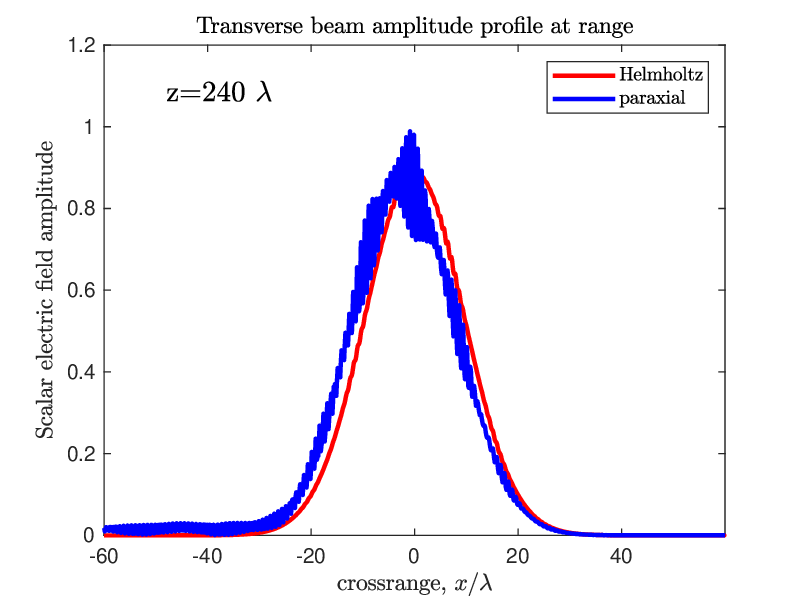}
      \caption{Beam amplitude profiles on the x-axis after $240 \lambda$ Helmholtz and
        paraxial propagations through a sine-product refractive index field
        with a $2 k_0 n_0 $ fundamental wavenumber}
      \label{Figure:sine_prod_n_xamp_5mu}
\end{figure}

\begin{figure}[ht]
      \includegraphics[width=.7\textwidth]{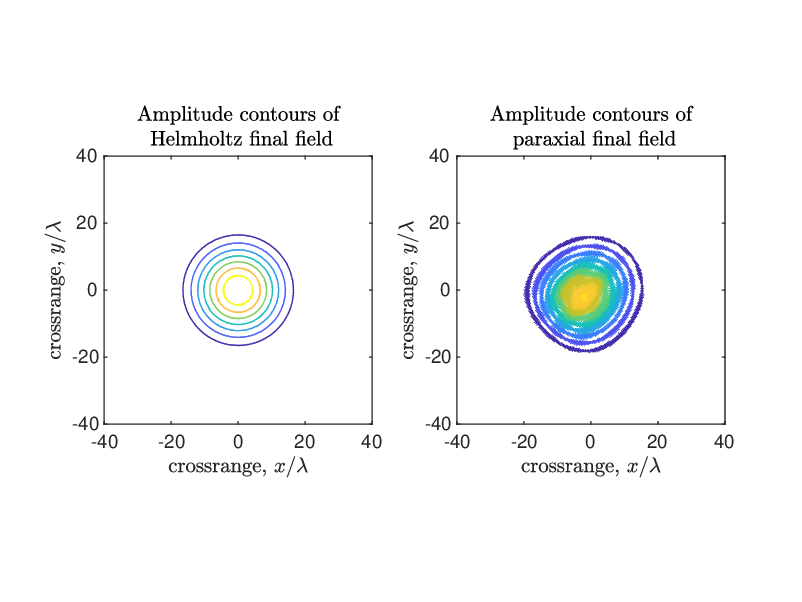}
      \caption{Transverse amplitude contours of Helmholtz and paraxial beams
        after a $z=240 \lambda$ propagation through a sine-product refractive
        index field with a $2 k_0 n_0$ fundamental wavelength}
      \label{Figure:sine_prod_n_amp_contour_5mu}
\end{figure}

Figures
\ref{Figure:sine_prod_n_xamp_20mu}, \ref{Figure:sine_prod_n_xamp_10mu}, and
\ref{Figure:sine_prod_n_xamp_5mu}
plot x-axis field amplitude profiles from a collimated Gaussian beam with ($1/e$) initial width of $25
\lambda$, after propagating $240 \lambda$ through each of the sine-product
refractive index fields. These are supplemented by Figures
\ref{Figure:sine_prod_n_amp_contour_10mu} and
\ref{Figure:sine_prod_n_amp_contour_5mu}, that show transverse amplitude
contours of the two beams after propagation through the two higher wavenumber
refractive index fields.  The
transverse physical domain of these calculations was $120 \lambda \times 120
\lambda$, with transverse grid sizes of $481 \times 481$ for the $.5 k_0
n_0$ and $k_0 n_0$ refractive index fields and $961 \times 961$ for the $2 k_0
n_0$ refractive index field. PML
boundary conditions of $60$ or $120$ points, respectively, completed the computational
domain.  Propagation space steps equalled one-quarter of the underlying
refractive index field wavelength.

For
the same refractive index field the beam amplitude response
calculated by the two propagators can be significantly different. The visible
effects displayed in these amplitude profiles and contours occur due to at least three
related phenomena: differing local response of the beam phase surface to
refractive index inhomogeneities; wholesale shifts of the beam centroid away
from the $(x,y)$ origin; and non-circular distortion of the beam in planes transverse
to the propagation direction.

For the $.5 k_0 n_0$ refractive index field results shown in Figure \ref{Figure:sine_prod_n_xamp_20mu}
there seems little to distinguish the Helmholtz and paraxial beams at
this propagation range.  However, a calculation of the beam centroids at the
end of the propagation shows that the paraxial centroid remains very close to
the $(x,y)$ origin at $(-0.00529, 0.00549) \lambda$, while the Helmholtz centroid has
shifted to $(0.217,-0.229) \lambda$. Transverse amplitude contour plots reveal
no perceptible non-circular distortion, and for that reason we omit them here. 

Stronger effects occur for propagations within the
$k_0 n_0$ and $2 k_0 n_0$ refractive index fields. Figure
\ref{Figure:sine_prod_n_xamp_10mu} shows the case when the refractive index field length
scale equals the fundamental wavelength in the Helmholtz equation.  Compared
to the paraxial beam, the
overall Helmholtz beam amplitude
decays more rapidly with propagation distance, and its local fluctuations
transverse to the beam are larger.  Transverse
amplitude contours are strongly distorted for the Helmholtz beam, but still
circular for the paraxial, as can be seen in Figure
\ref{Figure:sine_prod_n_amp_contour_10mu}. The Helmholtz beam
centroid has shifted to $(9.96, -11.36) \lambda$, while the paraxial centroid
remains close to the $(x,y)$ origin at $(0.0113,-0.050) \lambda$.

The two beams reverse their distortion and shift results for the highest
wavenumber refractive index field, shown in Figures \ref{Figure:sine_prod_n_xamp_5mu} and
\ref{Figure:sine_prod_n_amp_contour_5mu}.  The Helmholtz beam shows only the
smallest ripples in amplitude, while the paraxial beam amplitude fluctuates at 
levels that may exceed those of the Helmholtz beam in the $k_0 n_0$ refractive index
field. The Helmholtz beam centroid is very near the $(x,y)$ origin, $(-0.00007, 0.00005)
\lambda$, and its amplitude contours are circular.  The paraxial beam centroid, on the
other hand, has displaced to $(-4.74,-2.93) \lambda$, and its amplitude
contours are obviously non-circular, as well as seeming to be distorted in a
direction perpendicular to that suffered by the Helmholtz beam within the $k_0
n_0$ refractive index field.

The absence of amplitude fluctuations in the Helmholtz beam within the highest
wavenumber refractive index field relates to the evanescent cut-off that
occurs in Helmholtz, but not paraxial, propagation.  In the $2 k_0 n_0$
field the wave-like refractive index perturbations encountered by the Helmholtz beam are at
wavenumbers twice those that mark the evanescent boundary of the underlying
Helmholtz equation.  Thus the electric field phase perturbations they generate
are strongly
damped in the propagation direction. While continually
re-excited as the propagation advances, their effect decays in fractions of a
wavelength. The paraxial equation admits no such cut-off, so the refractive
index perturbations cause phase perturbations that continue to propagate as
undamped waves, interfering with previously excited waves at these frequencies, and
contributing to the high level of amplitude fluctuations seen in Figure 
\ref{Figure:sine_prod_n_xamp_5mu}.

\section{Discussion}

In this paper we have adapted analytical techniques pioneered in the
seismology and ocean acoustics communities to the problem of deriving a full
vector, one-way wave equation for electromagnetic wave propagation through
inhomogeneous refractive index fields.  The numerical technology used to
implement the exponential operator form of this
vector, one-way Helmholtz equation involves rational approximation, in
$\bar{\bm{Z}}$,  of the exponential operator \(\exp(i K \sqrt{1+\bar{\bm{Z}}})\),
followed by its partial fraction  decomposition. This
splits the original problem into a moderate number of independent
auxiliary problems whose results are summed at each propagation step to
advance the electric field in space. The $\bar{\bm{Z}}$ operator contains all
the propagation effects: diffraction, refraction, scattering, and electric field
component coupling.  Its discretization using high-resolving-efficiency finite
differences and PML boundary conditions turns each auxiliary problem into a
large, sparse, linear system that is iteratively solved by GMRES and conditioned by the
direct solve of a corresponding free space/constant-refractive-index
propagation problem that is memory efficient.  As an alternative, we have
derived the pseudo-differential form of the propagator along with a new
Cauchy-integral/partial fraction approximation of the
square root operator at its core.  In future work we expect to implement this
alternative form and test its performance and accuracy against the exponential operator form.
This new electromagnetic field propagation method is appropriate for
simulating both scalar and
vector wave problems involving high-wavenumber refractive index inhomogeneities, or
initial conditions with spatial support having wavelength or smaller scales.

\section*{Acknowledgements}

Approved for public release; distribution is unlimited. Public
Affairs release approval \# AFRL-2024-4623.  This work was supported by AFOSR grant
23RDCOR004.  L. Keefe and I. Zilberter
acknowledge support from the National Research
Council during part of this work.
The views expressed are those of the authors and do not necessarily reflect
the official policy or position of the Department of the Air Force, the
Department of Defense, or the U.S. Government.

\appendix
\section{}

\begin{figure}
      \includegraphics[width=0.5\textwidth]{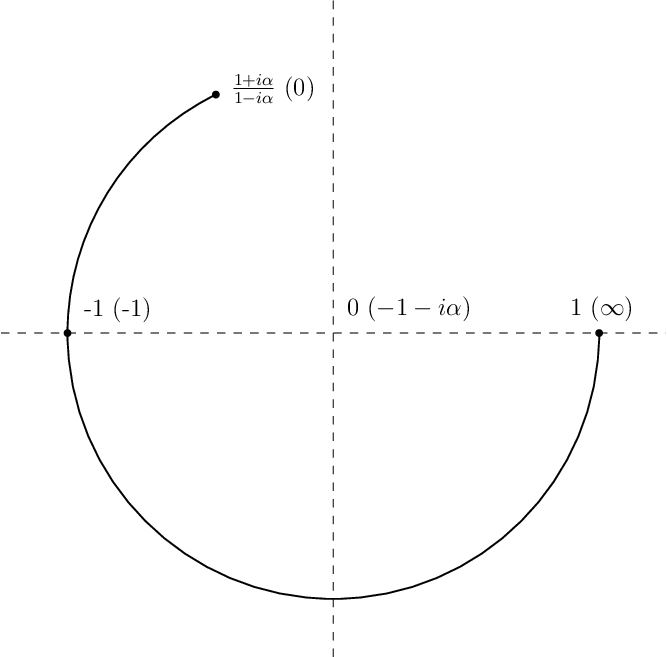}
      \caption{The $w$-plane corresponding to the M\"obius transformation given
          by equation~\eqref{eq:Mobius}. The bold circular arc is the image of the
          negative real line. For labeled points, the values in parentheses are the
          corresponding pre-image points in the $\zeta$-plane.}
       \label{Mobius_A}
\end{figure}

We complete the  Cauchy integral approximation of $f(\zeta) = (1 + \zeta)^{1/2}$ in this
section. The M\"obius tranformation \eqref{eq:Mobius} maps the real line into a
part of the unit circle.  The free parameter $\alpha > 0$ has the effect of moving the
point $\zeta = -1-i\alpha$ to the origin. The origin, in turn, is moved to the
point
$$ w_\alpha = \frac{1+i\alpha}{1-i\alpha} \; . $$
The segment $(-\infty,-1)$ is moved to the half
circle below the real axis and the segment $(-1,0)$ is moved to the circular arc
from $-1$ to $w_\alpha$ in the upper half plane. This geometry is shown in
Fig.~\ref{Mobius_A}.  As $\alpha$ goes to zero, the point $w_\alpha$ moves closer to
1, enlarging the arc in the upper half plane. Conversely, $\alpha$ going to
infinity moves the point $w_\alpha$ closer to -1, shrinking the
arc. It follows that choosing a smaller value of $\alpha$ will improve the
approximation near the branch point $\zeta = -1$.

\begin{figure}
      \includegraphics[width=0.5\textwidth]{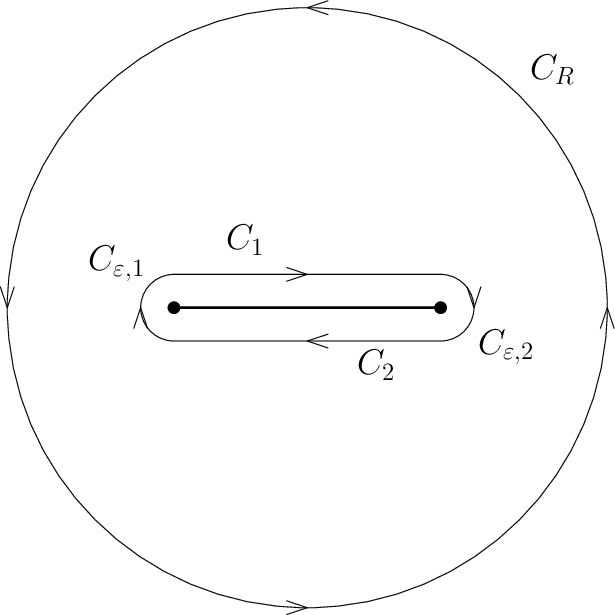}
      \caption{The contour used in the application of Cauchy's integral
          formula for the function $g(w)$. The bold line segment corresponds to the
          branch cut, from $[-1,1]$.}
      \label{Contour_A}
\end{figure}

To approximate the function
\begin{equation*}
  g(w) = f(\zeta(w)) = \sqrt{\alpha}e^{i\frac{\pi}4} (w+1)^{1/2} (w-1)^{-1/2} \; .
\end{equation*}
with the Cauchy integral formula \eqref{eq:Cauchy} we first define a new branch cut on $w \in [-1,1]$ using the local
coordinate parameterization
\begin{equation} \label{eq:LocalCoord_A}
  w+1 = r_1 e^{i\theta_1}, \quad w-1 = r_2 e^{i\theta_2}, \quad \theta_1,\, \theta_2 \in (-\pi,\pi) \; ,
\end{equation}
so that $g(w)$ is given by
\begin{equation} \label{eq:GBranch_A}
  g(w) = c_\alpha \sqrt{\frac{r_1}{r_2}} e^{i\frac{\theta_1-\theta_2}2}, \quad c_\alpha = \sqrt{\alpha} e^{i\frac{\pi}{4}} \; .
\end{equation}
For the Cauchy representation of $g$ to be valid, its integration contour must enclose a region
where $g$ is analytic. Take the contour shown in
Fig.~\ref{Contour_A}, so that
\begin{equation*}
  g(w) = \frac{1}{2\pi i} \left( \int_{C_R} + \int_{C_{\varepsilon,1}} + \int_{C_1} + \int_{C_{\varepsilon,2}} + \int_{C_2} \right) \frac {g(\xi)}{\xi - w} d\xi \; ,
\end{equation*}
where $C_R$ is the circle of radius $R$ centered at the origin and $C_{\varepsilon,1}$ and
$C_{\varepsilon,2}$ are semicircles around -1 and 1, respectively, of radius
$\varepsilon$. We are interested in the limit as $R \rightarrow \infty$ and
$\varepsilon \rightarrow 0$.

On $C_R$ we have $\xi = Re^{i\theta}$, $\theta \in (-\pi,\pi)$. Using the
local coordinates defined in \eqref{eq:LocalCoord_A}, we have $\xi+1
\sim Re^{i\theta}$ and $\xi-1 \sim Re^{i\theta}$. From 
\eqref{eq:GBranch_A} the function $g$ on the contour is given by
$g(\xi) \sim c_\alpha R^{1/2} e^{i\theta/2} R^{-1/2} e^{-i\theta/2} =
c_\alpha$.  Therefore,
\begin{align*} 
  \int_{C_R} \frac {g(\xi)}{\xi - w} d\xi &\sim c_\alpha \int_{-\pi}^{\pi} \frac{1}{Re^{i\theta}-w} \left( iRe^{i\theta}d\theta \right) \\
                                                &\sim c_\alpha i\int_{-\pi}^{\pi} d\theta \\
                                                &= 2\pi c_\alpha i \; .
\end{align*} 

On $C_{\varepsilon,2}$ we have $\xi = 1 + \varepsilon e^{i\theta}$, $\theta
\in (\pi/2, -\pi/2)$, so that $\xi+1 \sim 2$ and $\xi-1 = \varepsilon
e^{i\theta}$, implying $g(\xi) \sim c_\alpha \sqrt{\frac 2{\varepsilon}}
e^{-i\theta/2}$. Therefore,
\begin{align*} 
  \int_{C_{\varepsilon,2}} \frac {g(\xi)}{\xi - w} d\xi &\sim c_\alpha \int_{\frac \pi 2}^{-\frac \pi 2} \frac{\sqrt{\frac 2{\varepsilon}} e^{-i\theta/2}}{1 + \varepsilon e^{i\theta}-w} \left( i\varepsilon e^{i\theta} d\theta \right) \\
                                                &\sim -i c_\alpha \sqrt{2\varepsilon} \int_{-\frac \pi 2}^{\frac \pi 2} \frac{e^{i\theta/2}}{1 + \varepsilon e^{i\theta}-w} d\theta \\
                                                &\rightarrow 0 \; .
\end{align*} 
Similarly, $\int_{C_{\varepsilon,1}} \rightarrow 0$.

On $C_1$ as $\varepsilon \rightarrow 0$ we have $\xi = \eta$ where $\eta$
goes from -1 to 1, $\theta_1 = 0$, and $\theta_2 = \pi$, implying $g(\xi) = c_\alpha
\sqrt{\frac{1+\eta}{1-\eta}} e^{-i\pi/2} = -i c_\alpha \sqrt{\frac{1+\eta}{1-\eta}}$.
Therefore,
\begin{align*} 
  \int_{C_1} \frac{g(\xi)}{\xi - w} d\xi &= -i c_\alpha \int_{-1}^{1} \frac 1{\eta-w} \sqrt{\frac{1+\eta}{1-\eta}} d\eta \; .
\end{align*} 

On $C_2$ as $\varepsilon \rightarrow 0$ we have $\xi = \eta$ where $\eta$
goes from 1 down to -1, $\theta_1 = 0$, and $\theta_2 = -\pi$, implying $g(\xi) = c_\alpha
\sqrt{\frac{1+\eta}{1-\eta}} e^{i\pi/2} = i c_\alpha \sqrt{\frac{1+\eta}{1-\eta}}$.
Therefore,
\begin{align*} 
  \int_{C_2} \frac{g(\xi)}{\xi - w} d\xi &= i c_\alpha \int_{1}^{-1} \frac 1{\eta-w} \sqrt{\frac{1+\eta}{1-\eta}} d\eta \\
                                               &= -i c_\alpha \int_{-1}^{1} \frac 1{\eta-w} \sqrt{\frac{1+\eta}{1-\eta}} d\eta \; .
\end{align*} 

Combining results, as $R \rightarrow \infty$ and $\varepsilon \rightarrow 0$, we have
\begin{equation}
  g(w) = c_\alpha \left( 1 - \frac 1\pi \int_{-1}^{1} \frac 1{\eta-w} \sqrt{\frac{1+\eta}{1-\eta}} d\eta \right) .
\end{equation}
%


\bibliography{ehelm_prop.bib}

\end{document}